\documentclass{emulateapj}

\begin{document}

\shortauthors{Luhman et al.}
\shorttitle{Infrared/X-ray Survey of Taurus}

\title{An Infrared/X-ray Survey for New Members of the Taurus Star-Forming 
Region\altaffilmark{1}}

\author{
K. L. Luhman\altaffilmark{2,3},
E. E. Mamajek\altaffilmark{4},
P. R. Allen\altaffilmark{2},
and K. L. Cruz\altaffilmark{5,6}
}

\altaffiltext{1}
{Based on observations performed with the Hobby-Eberly Telescope,
the Magellan Telescopes at Las Campanas Observatory, the NASA Infrared
Telescope Facility, Gemini Observatory, the {\it Spitzer Space Telescope}, the
{\it XMM-Newton Observatory}, and the Canada-France-Hawaii Telescope. 
}

\altaffiltext{2}{Department of Astronomy and Astrophysics, The Pennsylvania
State University, University Park, PA 16802; kluhman@astro.psu.edu.}

\altaffiltext{3}{Visiting Astronomer at the Infrared Telescope Facility, which
is operated by the University of Hawaii under Cooperative Agreement no.\ NCC
5-538 with the National Aeronautics and Space Administration (NASA),
Office of Space Science, Planetary Astronomy Program.}

\altaffiltext{4}{Department of Physics and Astronomy,
The University of Rochester, Rochester, NY 14627.}

\altaffiltext{5}{Astronomy Department, California Institute of Technology, 
Pasadena, CA 91125.}

\altaffiltext{6}{{\it Spitzer} Fellow.}

\begin{abstract}

We present the results of a search for new members of the Taurus
star-forming region using data from the {\it Spitzer Space Telescope}
and the {\it XMM-Newton Observatory}. We have obtained optical and
near-infrared spectra of 44 sources that exhibit red {\it Spitzer}
colors that are indicative of stars with circumstellar disks and 51
candidate young stars that were identified by Scelsi and coworkers
using {\it XMM-Newton}. We also performed spectroscopy on four
possible companions to members of Taurus that were reported by Kraus
and Hillenbrand.  Through these spectra, we have demonstrated the
youth and membership of 41 sources, 10 of which were independently
confirmed as young stars by Scelsi and coworkers.  Five of the new
Taurus members are likely to be brown dwarfs based on their late
spectral types ($>$M6).  One of the brown dwarfs has a spectral type
of L0, making it the first known L-type member of Taurus and the least
massive known member of the region ($M\sim4$-7~$M_{\rm Jup}$). Another
brown dwarf exhibits a flat infrared spectral energy distribution,
which indicates that it could be in the protostellar class~I stage
(star+disk+envelope).  Upon inspection of archival images from various
observatories, we find that one of the new young stars has a large
edge-on disk ($r=2\farcs5=350$~AU).  The scattered light from this
disk has undergone significant variability on a time scale of days in
optical images from the Canada-France-Hawaii Telescope.  Using the
updated census of Taurus, we have measured the initial mass function
for the fields observed by {\it XMM-Newton}. The resulting mass
function is similar to previous ones that we have reported for Taurus,
showing a surplus of stars at spectral types of K7-M1
(0.6-0.8~$M_\odot$) relative to other nearby star-forming regions like
IC~348, Chamaeleon~I, and the Orion Nebula Cluster.
\end{abstract}

\keywords{accretion, accretion disks --- planetary systems: protoplanetary 
disks --- stars: formation --- stars: low-mass, brown dwarfs --- 
stars: luminosity function, mass function}

\section{Introduction}
\label{sec:intro}

The Taurus complex of dark clouds is one of the best sites for
studying the formation of stars in a quiescent, relatively isolated
environment.  It is among the nearest star-forming regions
($d=140$~pc) and exhibits a very low stellar density
($n\sim1$-10~pc$^{-3}$).  Although the individual clouds are sparsely
populated, the cloud complex as a whole contains more than 300 known
members.  Working toward a complete census of Taurus is important for
the identification of rare objects (e.g., edge-on disks, transitional
disks, protostars) as well as the statistical characterization of the
stellar population (e.g., disk fraction, initial mass function,
spatial distribution). A variety of methods have been employed in
surveys for new members of Taurus
\citep{ken08}. Two of these techniques, mid-infrared (IR) imaging and X-ray 
imaging, are highly complementary.
Mid-IR observations can identify stars that have circumstellar disks 
and can penetrate the high levels of extinction that surround stars at
the earliest evolutionary stages while X-ray data 
can uncover the diskless members of a young stellar population.

Because of their excellent sensitivities and large fields of view, the
{\it Spitzer Space Telescope} \citep{wer04} and the {\it XMM-Newton
Observatory} \citep{jan01} are the best available telescopes for
wide-field imaging surveys at mid-IR and X-ray wavelengths,
respectively. The unique capabilities of these facilities have been
applied to the Taurus star-forming region through the Taurus {\it
Spitzer} Legacy Survey (D. Padgett, in preparation) and the {\it
XMM-Newton} Extended Survey of the Taurus Molecular Cloud
\citep[XEST,][]{gud07}.
\citet{luh06tau2,luh09fu} and \citet{sce07,sce08} have used the data
from these surveys to search for new members of Taurus. We have
continued those efforts by performing spectroscopy on IR sources that
we have identified in the {\it Spitzer} images and X-ray sources that
were reported by \citet{sce07}. In this paper, we describe the
selection of these candidate members of Taurus (\S~\ref{sec:select})
and measure their spectral types with optical and IR spectra
(\S~\ref{sec:spec}). We then characterize the stellar parameters of
the new members and discuss other notable properties of these objects
(\S~\ref{sec:prop}). Finally, we use our updated census of the stellar
population in Taurus to measure the initial mass function (IMF) within
the fields observed by XEST (\S~\ref{sec:imf}).

\section{Selection of Candidate Members of Taurus}
\label{sec:select}

For the IR selection of candidate members of Taurus, we used images at
3.6, 4.5, 5.8, and 8.0~\micron\ obtained with {\it Spitzer}'s Infrared
Array Camera \citep[IRAC;][]{faz04} and images at 24~\micron\ obtained
with the Multiband Imaging Photometer for {\it Spitzer}
\citep[MIPS;][]{rie04}.  We considered all observations of this kind
that have been performed in Taurus, many of which were collected
through the {\it Spitzer} Legacy program of D.\ Padgett. These images
of Taurus encompass a total area of 46~deg$^{2}$.  The characteristics
of the individual IRAC and MIPS observations are summarized by
\citet{luh09tau}, who present a compilation of 3.6-24~\micron\ {\it
Spitzer} photometry for all known members of Taurus.  To identify
candidate members of Taurus, we searched the {\it Spitzer} images for
stars exhibiting red IRAC colors that are indicative of excess
emission from circumstellar disks and envelopes. The reduction and
analysis methods were the same as those employed in our previous
surveys of star-forming regions \citep[][references
therein]{luh08cha2}.  In addition to sources with IRAC excesses, we
also inspected the data for stars with excess emission at 24~\micron\
but not in the IRAC bands, which is a signature of a disk with an
inner hole. We selected 44 of the resulting candidates for followup
spectroscopy to determine whether they are members of Taurus. Two of
these sources, FU~Tau~A and B, were reported in a separate study
\citep{luh09fu}. Because the secondary is too close to the primary for the
measurement of IRAC photometry through our automated procedures, it
was not identified as a candidate based on its IRAC colors. Instead,
FU~Tau~B was selected for spectroscopy because of its close proximity
to FU~Tau~A. Nevertheless, we count the former among the IR candidates
for the purposes of this study.  Through the spectroscopic
observations described in the next section, we find that 24 of the 44
candidates are members of Taurus. One of these new members was
detected by {\it XMM-Newton} (XEST~26-071) but was not recognized as a
candidate member with those data \citep{sce07}.

\begin{figure}
\epsscale{1}
\plotone{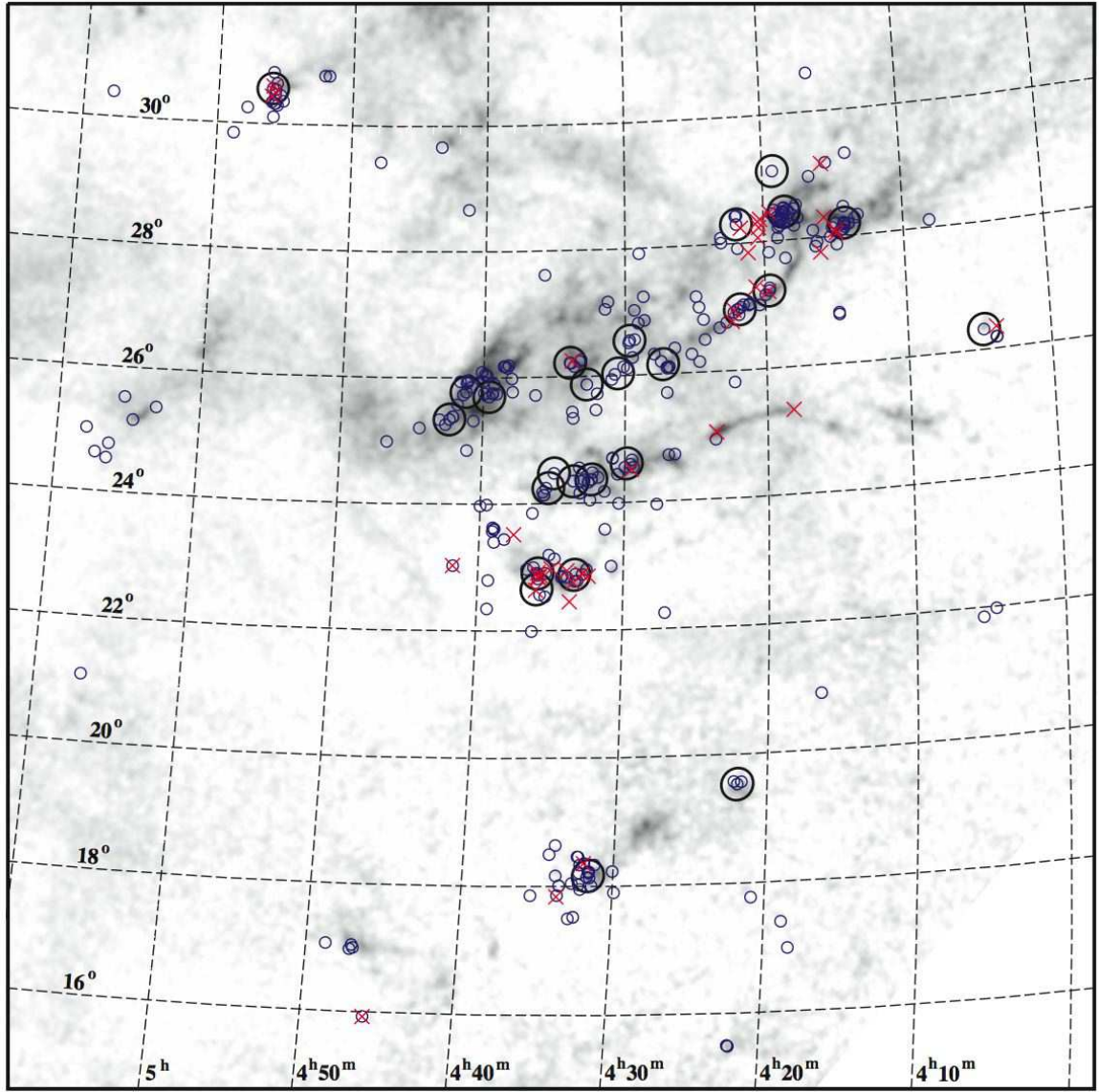}
\caption{
Spatial distribution of previously known members of the Taurus
star-forming region ({\it small circles}) and new objects that we have 
confirmed as members through spectroscopy ({\it crosses}).
The latter were identified as possible members with the 
{\it XMM-Newton} Observatory \citep{sce07}, the {\it Spitzer Space Telescope}
(this work), and a companion survey with 2MASS \citep{kra07}.
The fields imaged with {\it XMM-Newton} are indicated ({\it large circles}).
The dark clouds in Taurus are displayed with a map of extinction 
\citep[{\it grayscale},][]{dob05}.
}
\label{fig:map}
\end{figure}

We have included in our spectroscopic sample candidate members of
Taurus that have been found through X-ray observations by the {\it
XMM-Newton Observatory}. The XEST program \citep{gud07} obtained
images of 19 fields in Taurus with {\it XMM-Newton} and utilized
archival data for seven additional fields (one of which was observed
twice). The boundaries of the XEST images are indicated on the map of
the Taurus cloud complex in Figure~\ref{fig:map}. These fields are
primarily located in the densest stellar aggregates and encompass a
total area of 5~deg$^{2}$.  Using these data, \citet{sce07} identified
57 possible members of Taurus.  One of these candidates, XEST~13-010,
was reported as a member by
\citet{luh06tau2} while another candidate, XEST~06-045, is a galaxy according 
to images from the Digitized Sky Survey (DSS). We excluded from
consideration four candidates that are far from the Taurus clouds
($\alpha<4$~hours).  We selected the remaining 51 candidates for
spectroscopy, 16 of which are classified as members of Taurus in the
next section.

In addition to the IR and X-ray candidates, we have performed
spectroscopy on four possible companions to known members of Taurus
from \citet{kra07}, one of which is classified as a Taurus member
through our spectroscopy. We also observed a previously known but
widely overlooked member, LH~0429+17 \citep{rei99}, so that we could
measure its spectral type with the same classification methods that we
have been applied to the other late-type members of Taurus.

\section{Spectroscopy of Candidates}
\label{sec:spec}

\subsection{Spectral Classification}

We performed optical and near-IR spectroscopy on the 100 targets
selected in the previous section using a variety of instruments and
telescopes. The dates, telescopes, and instrument configurations for
these observations are summarized in Table~\ref{tab:log}.  We examined
these data for signatures of youth that indicate membership in Taurus,
as done in our previous studies of this kind
\citep[e.g.,][]{luh04cha}.  In addition, we considered other diagnostics 
of membership when available, such as proper motions 
\citep[][\S~\ref{sec:app}]{luh09fu}.
Through this analysis, we classified 42 of the 100 targets in our spectroscopic
sample as members of Taurus. The evidence of youth and membership for
these sources is compiled in Table~\ref{tab:new}. The 58 nonmembers
are listed in Table~\ref{tab:non}.

\begin{figure}
\epsscale{0.7}
\plotone{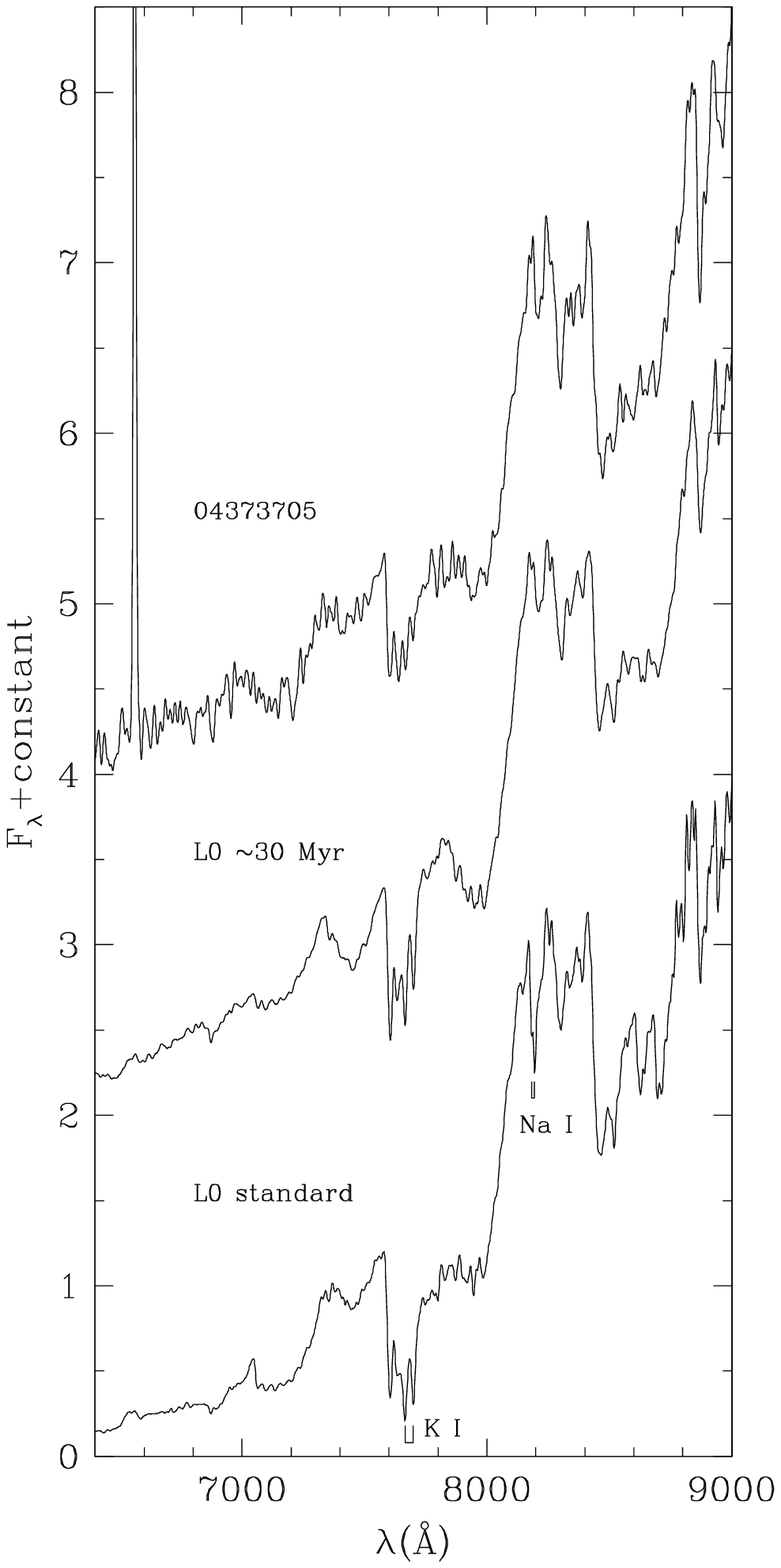}
\caption{
Optical spectrum of the new Taurus member 2MASS~J04373705+2331080 compared 
to data for the young L0 dwarf 2MASS~01415823-4633574 \citep{kir06,cru09}
and the L0 standard 2MASS~J03454316+2540233 \citep{kir99}.
The proper motion of 2MASS~01415823-4633574 from \citet{fah09} is 
consistent with membership in the Tucana-Horologium association, 
which has an age of $\sim30$~Myr \citep{tor00,zuc04}.
The data are displayed at a resolution of 13~\AA\ and are normalized at
7500~\AA.
}
\label{fig:op3}
\end{figure}

\begin{figure}
\epsscale{1.1}
\plotone{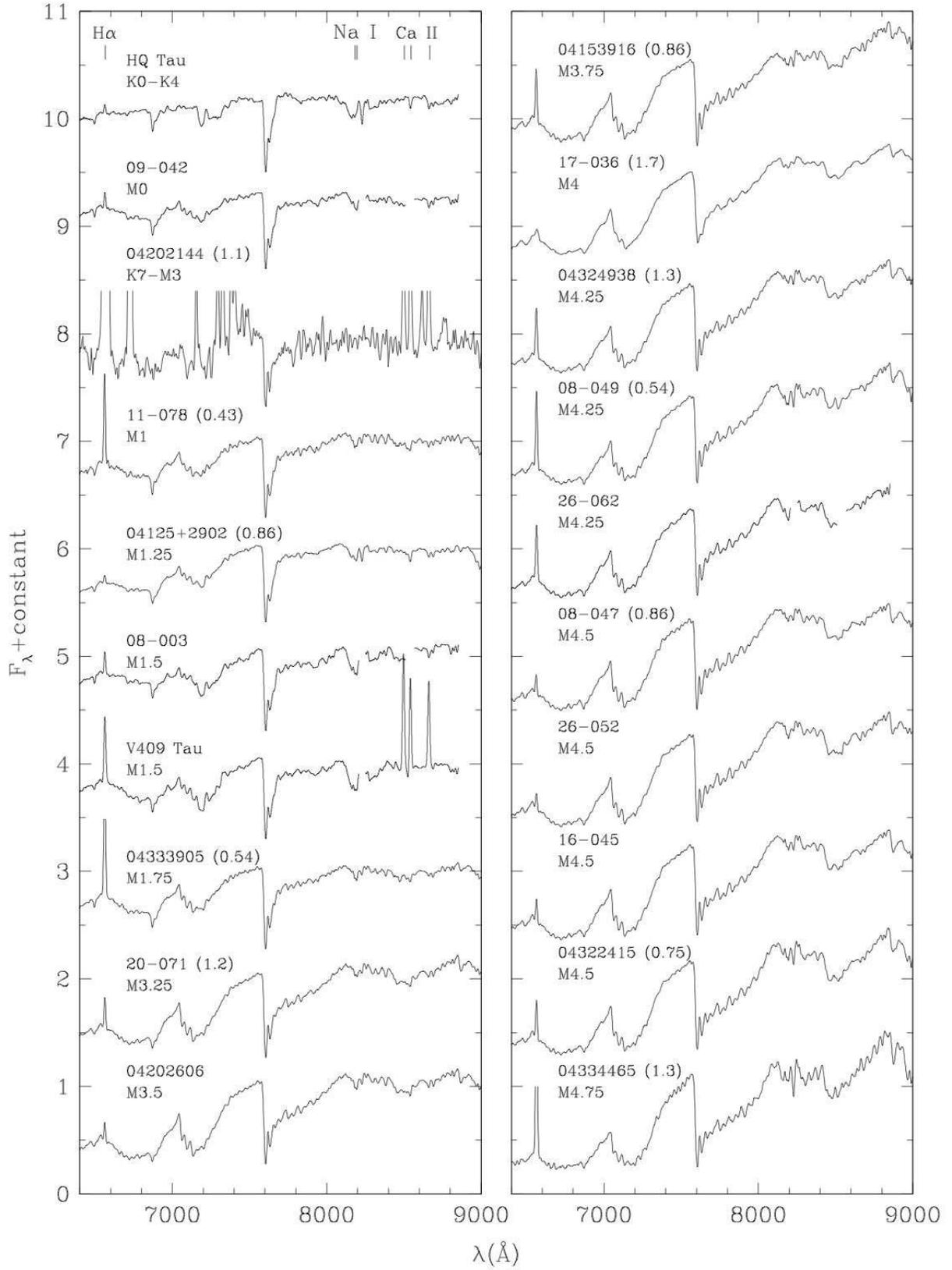}
\caption{
Optical spectra of new members of Taurus.
The spectra have been corrected for extinction, which is quantified in
parentheses by the magnitude difference of the reddening
between 0.6 and 0.9~\micron\ ($E(0.6-0.9)$).
The data are displayed at a resolution of 13~\AA\ and are normalized at
7500~\AA.
}
\label{fig:op1}
\end{figure}

\begin{figure}
\epsscale{1.1}
\plotone{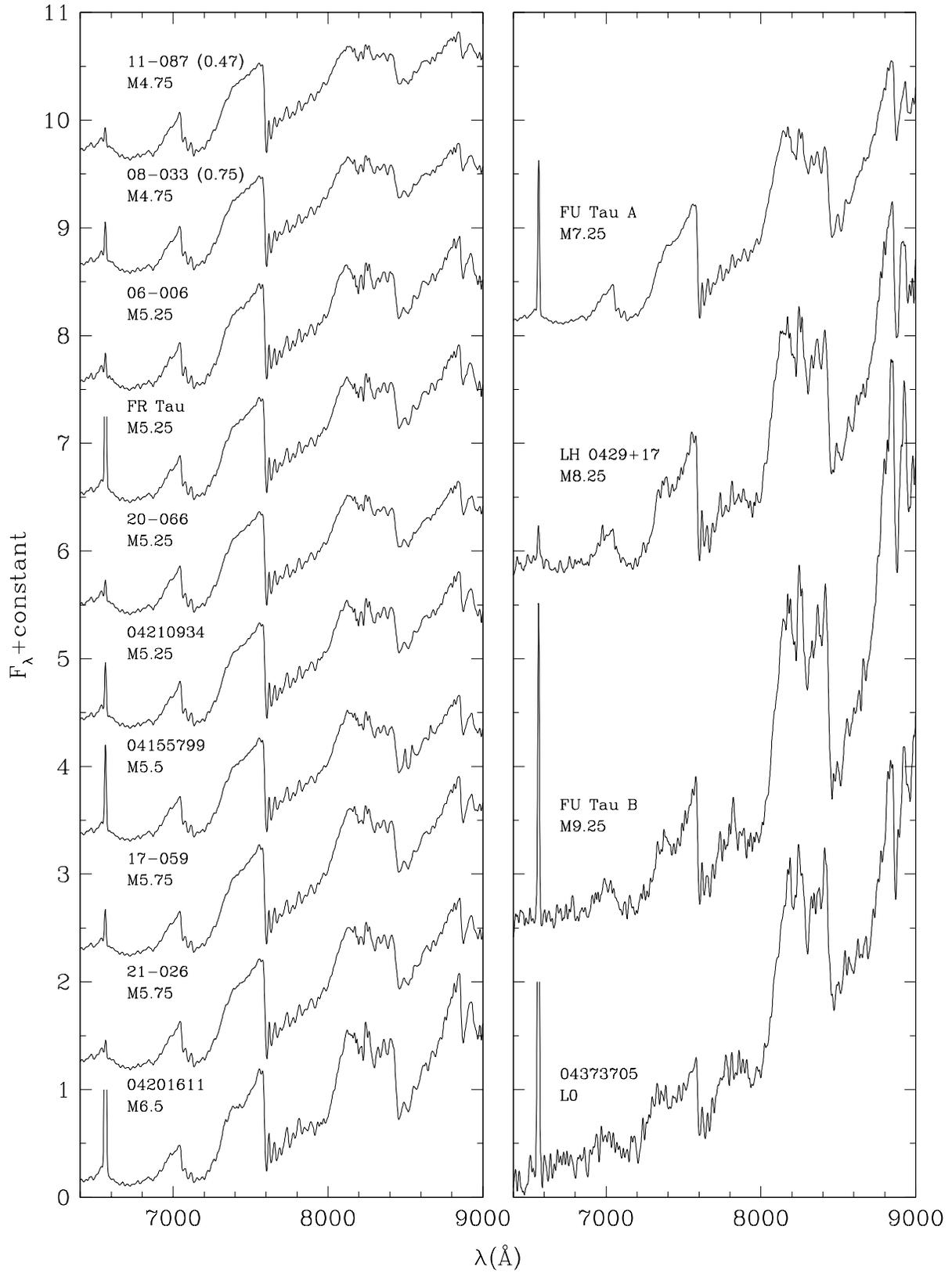}
\caption{
More optical spectra of new members of Taurus (see Fig.~\ref{fig:op1})
and the previously known member LH~0429+17 \citep{rei99}.
}
\label{fig:op2}
\end{figure}

\begin{figure}
\epsscale{1.1}
\plotone{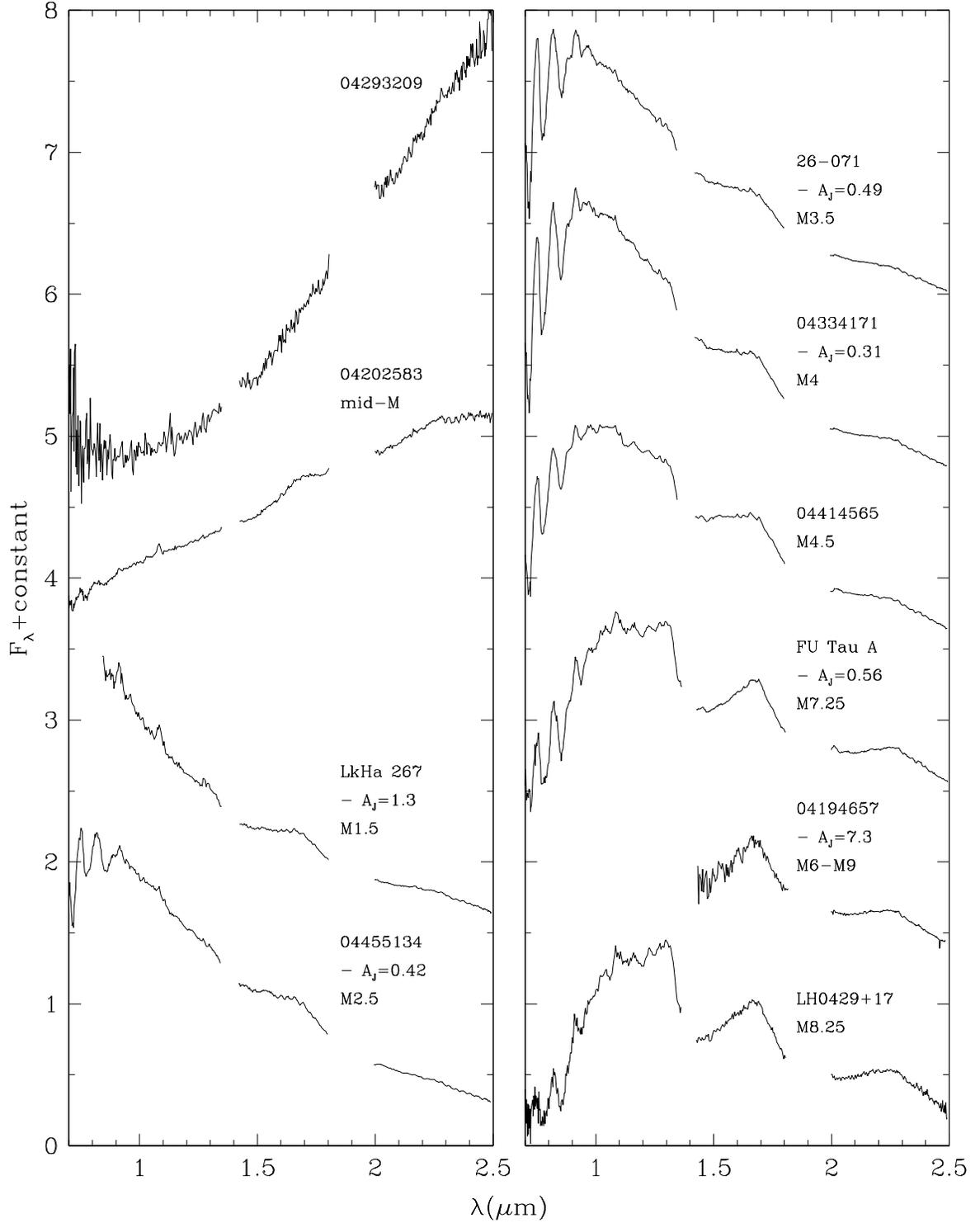}
\caption{
SpeX near-IR spectra of new members of Taurus and the previously known 
member LH~0429+17 \citep{rei99}. 
FU~Tau~A and LH~0429+17 are labeled with spectral types measured
from the optical spectra in Figure~\ref{fig:op2}. The remaining sources
have been classified with these IR data.
Most of the spectra have been dereddened to match the slopes
of the young optical standards. For the first two sources, we show the 
observed data without dereddening because their classifications are uncertain. 
These data have a resolution of $R=100$ and are normalized at 1.68~\micron.
}
\label{fig:ir}
\end{figure}

\begin{figure}
\epsscale{0.6}
\plotone{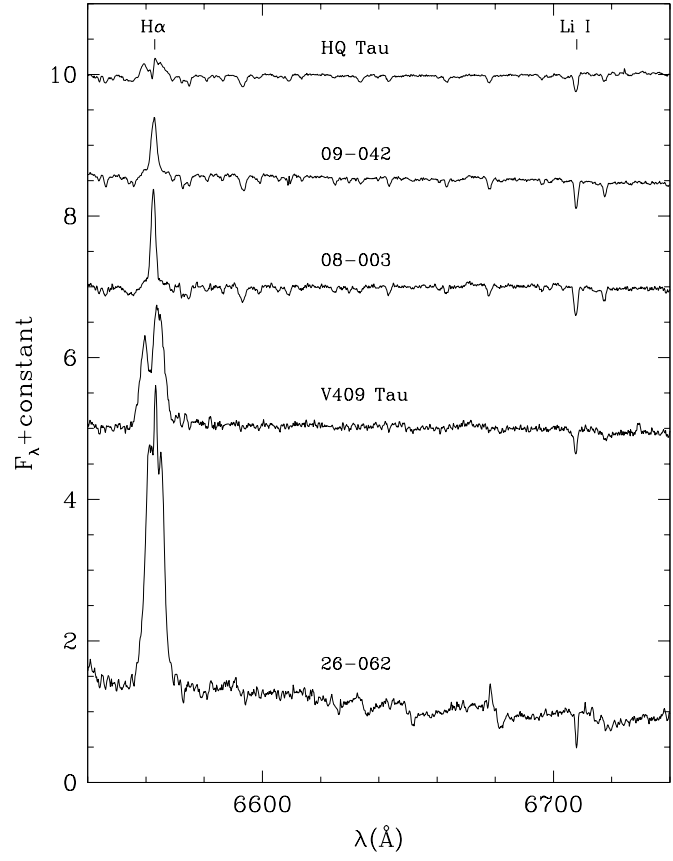}
\caption{
High-resolution spectra of H$\alpha$ and Li~I for new members of Taurus.
The data have a resolution of 0.7~\AA\ and are normalized to the 
continuum near the Li~I line. 
}
\label{fig:li}
\end{figure}

Most of the members of Taurus in our sample exhibit late-type ($>$M0)
features in their spectra (H$_2$O, TiO, VO). To measure spectral types
for these sources, we have compared their spectra to previous data
that we have collected for known members of Taurus and other
star-forming regions
\citep{luh04cha,luh04tau,luh06tau1}, which were originally classified at 
optical wavelengths through comparison to averages of dwarfs and giants 
\citep{luh99}. One new member, 2MASS~J04373705+2331080 (hereafter 
2M~0437+2331), is later than all previously known members of Taurus
($\leq$M9.5). We classified this object through a comparison to
standard L dwarfs \citep{kir99} and young L dwarfs in the field
\citep[$\tau\sim10$-100~Myr,][]{kir06,cru09}, arriving at a spectral 
type of L0 (see Figure~\ref{fig:op3}). The one new member with a
K-type spectrum, HQ~Tau, was classified with dwarf standards
\citep{as95}.  We could not measure a spectral type for one of the new
members, 2MASS~J04293209+2430597, because photospheric features are
not detected in its spectrum. Our spectral classifications for the
Taurus members are provided in Table~\ref{tab:new}. The spectra are
shown in order of spectral type in
Figures~\ref{fig:op1}-\ref{fig:ir}. The highest resolution data are
presented in Figure~\ref{fig:li} for the wavelength range encompassing
H$\alpha$ and Li~I. The equivalent widths of Li~I from these spectra
are given in Table~\ref{tab:li}. The positions of the 41 new members
(excluding LH~0429+17) are plotted on the map of Taurus in
Figure~\ref{fig:map}.

Among the nonmembers, we classified field stars with standard dwarfs
and giants \citep{hen94,kir91,kir97,cus05} and we identified galaxies
based on the presence of redshifted emission lines.  The
classifications of the nonmembers are found in Table~\ref{tab:non}.

\citet{sce08} obtained optical spectra of 25 of the candidate members
of Taurus from the XEST program \citep{sce07}.  They classified 10
candidates as young stars and 12 candidates as nonmembers.  We
observed 20 of these 22 sources; our membership classifications agree
with those of \citet{sce08} in all cases.  The membership of the
remaining 3 candidates was uncertain based on the data from
\citet{sce08}. We find that two of these objects are foreground dwarfs
(XEST~08-014, XEST~15-034) while the other candidate is a Taurus
member (XEST~20-071).  Our spectral types are systematically later
than those from \citet{sce08} by a few subclasses for sources in
common between the two studies.

\subsection{Comments on Individual Sources}
\label{sec:comm}

Several of our targets have displayed evidence of membership in Taurus
in previous studies, although they lacked spectroscopic
classifications.  The previous observations of one of these objects,
FU~Tau, are described in detail by \citet{luh09fu}.
\citet{tor95} identified 2MASS~J04455134+1555367 as a possible young star based
on mid-IR photometry from the {\it Infrared Astronomical Satellite}
(IRAS)\footnote{2MASS~J04455134+1555367 and its $20\arcsec$ companion
HD~30171 were unresolved in the IRAS data. Images at higher resolution
from {\it Spitzer} demonstrate that 2MASS~J04455134+1555367 was the
source of the IRAS emission \citep{luh09tau}.}.  They detected
emission in H$\alpha$ and He~I and absorption in Li~I through followup
spectroscopy.
\citet{jh79} found that the proper motion of HQ~Tau is consistent 
with membership in Taurus. Mid-IR excess emission also was detected
toward this star in photometry from IRAS \citep{har88} and a spectrum
from {\it Spitzer} \citep{fur06}.
\citet{ken94} identified V409~Tau and IRAS~04125+2902 as candidate
members of Taurus based on IRAS data. They classified the latter as a
galaxy through near-IR images or optical spectroscopy, but it is an
M-type star according to our spectroscopy (see Figure~\ref{fig:op1}).

We now discuss the stars in our sample that have uncertain
classifications.  Our spectrum of 2MASS~J04293209+2430597 does not
show any absorption or emission lines that would demonstrate that it
is a young star. However, its very red, featureless spectrum and
mid-IR excess emission are consistent with a protostar. Given its
close proximity to a dark cloud and other known members of Taurus, we
tentatively classify it as a member. Based on its strong H$\alpha$
emission [$W_{\lambda}$(H$\alpha)=58\pm1$~\AA] and mid-IR excess
emission, 2MASS~J04124858+2749563 is clearly a young star.  However,
it is much fainter than members of Taurus near its spectral type and
its proper motion is inconsistent with membership
\citep[][\S~\ref{sec:app}]{mon03,zac04b}. Therefore, 
2MASS~J04124858+2749563 is probably a background star (albeit a young
one). Our IR spectrum of 2MASS~J04345973+2807017 is better matched by
dwarf standards, but it does not have a sufficient signal-to-noise
ratio to rule out youth. This star is anomalously faint for a Taurus
member and is far from known members of the region, which suggest that
it is a background star.  Stars with edge-on disks also appear very
faint for their spectral types.  2MASS~J04345973+2807017 does appear
to have excess emission at 8~\micron\ that indicates the presence of a
disk, but its large distance from known members of Taurus tends to
support a classification as a background star. Finally, the proper
motion of the A-type star 2MASS~J04180338+2440096 is inconsistent with
membership in Taurus \citep[][\S~\ref{sec:app}]{hog00,zac04b,ros08}, but it 
exhibits mid-IR excesses of 0.2 and 3~mag at 8 and 24~\micron,
respectively.  It is probably a field star with a debris disk.

\section{Properties of New Members}
\label{sec:prop}

\subsection{Temperatures and Luminosities}
\label{sec:hr}

We have estimated the effective temperatures and bolometric
luminosities of the Taurus members in our spectroscopic sample so that
we can place them on the Hertzsprung-Russell (H-R) diagram. During
spectral classification, we measured extinctions from the slopes of
most of our optical and near-IR spectra
\citep{luh04cha,luh07cha}. We could not use the MRS spectra for this purpose
because they were obtained through fibers rather than slits aligned at
the parallactic angle, making the data susceptible to differential refraction
and the spectral slopes unreliable. For the MRS targets, we computed 
extinctions from $J-H$ and $H-K_s$ in the manner described by \citet{luh04cha}.
We estimated luminosities by combining the extinctions with 
$J$-band photometry from the Point Source Catalog of the
Two-Micron All-Sky Survey \citep[2MASS;][]{skr06}, a distance of 140~pc 
\citep{wic98,loi05,loi07,tor07,tor09}, and the bolometric corrections used
by \citet{luh04cha,luh07cha}. 
The luminosities of 2MASS~J04194657+2712552 (hereafter 2M~0419+2712) 
and FU~Tau~B are based on $H$ and $K_s$, respectively, since reliable 
$J$-band measurements are unavailable. 
We have converted our spectral types to effective temperatures with the
temperature scales from \citet{sk82} and \citet{luh03ic} for $<$M0 and
$\geq$M0, respectively. As done by \citet{luh08cha1},
we adopt a temperature of 2200~K for L0. 
The extinctions, effective temperatures, and bolometric luminosities 
are presented in Table~\ref{tab:new}.
We cannot measure these parameters for 2MASS~J04293209+2430597 
because it lacks a spectral classification.

For comparison to the new Taurus members on the H-R diagram, we have
compiled temperatures and luminosities for previously known members
(\S~\ref{sec:app}). We exclude members that are anomalously faint for
their spectral types because their luminosity estimates are probably
unreliable. We treat multiple systems that are unresolved by 2MASS as
single objects.  We adopt extinctions that have been derived from our
previous optical and near-IR spectroscopy
\citep{luh03tau,luh04tau,luh06tau1,luh06tau2}
and additional unpublished 0.8-2.5~\micron\ spectra obtained with SpeX. 
For members that lack spectroscopic data of this kind, 
we estimate extinctions from $J-H$ and $H-K_s$. 
Luminosities are based on measurements of $J$ from 2MASS when available.
We adopt $J$ from multiplicity surveys for a few systems that are 
marginally resolved by 2MASS.

\begin{figure}
\epsscale{1}
\plotone{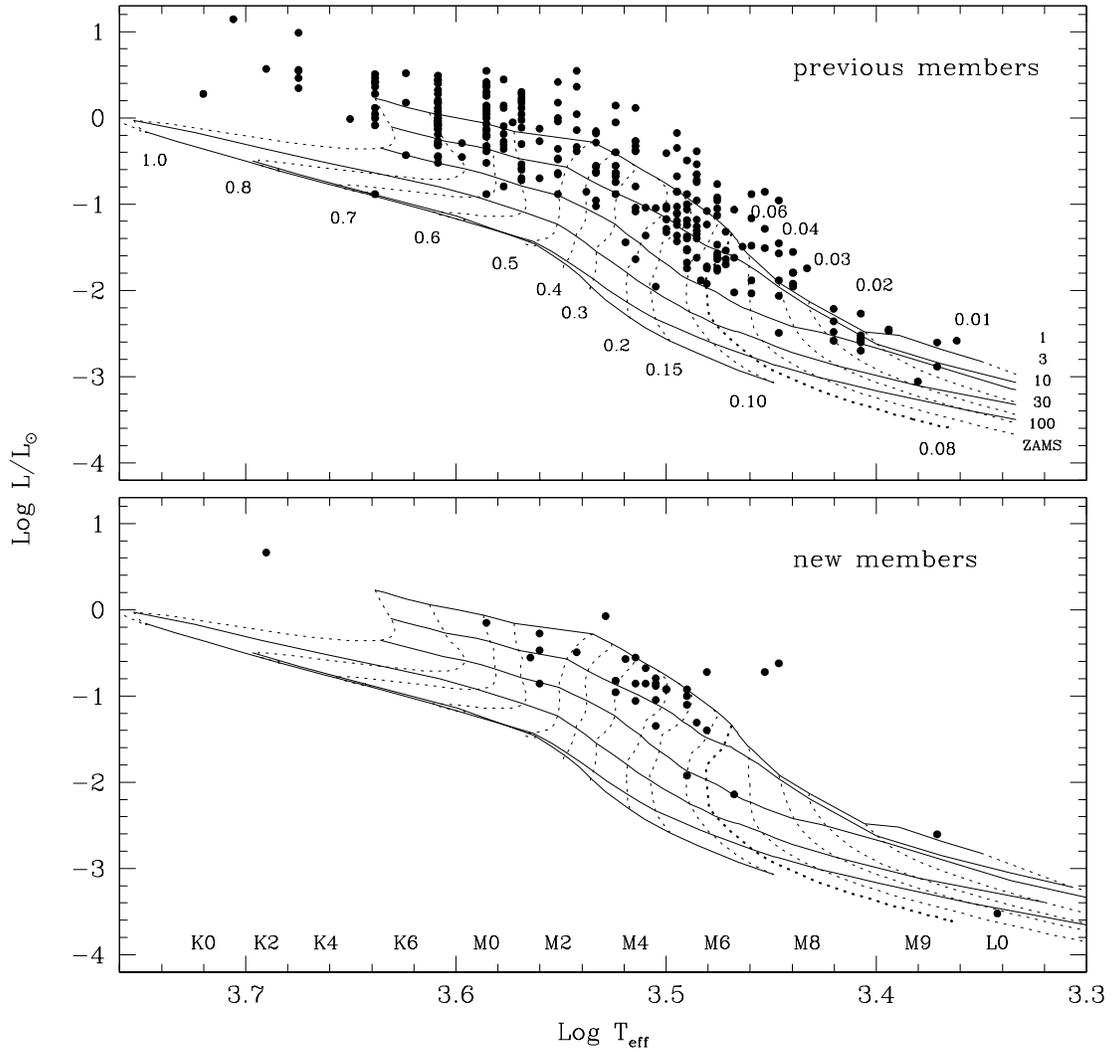}
\caption{
H-R diagram for previously known members of Taurus ({\it top}) and new members
discovered by \citet{sce07,sce08} and in this work ({\it bottom}).
We have omitted companions that are unresolved by
2MASS and stars that appear on or below the main sequence. 
The anomalously low luminosity estimates for several of those stars are known 
to result from edge-on disks (e.g., Figure~\ref{fig:edge}). 
These data are shown with the theoretical evolutionary models of
\citet{bar98} ($0.1<M/M_\odot\leq1$) and \citet{cha00} ($M/M_\odot\leq0.1$),
where the mass tracks ({\it dotted lines}) and isochrones ({\it solid lines})
are labeled in units of $M_\odot$ and Myr, respectively.
}
\label{fig:hr}
\end{figure}

Our luminosity estimates for the new members XEST~11-078, 2MASS
J04202583+2819237, 2MASS J04333905+2227207, and 2MASS
J04202144+2813491 would place these stars well below the sequence of
known Taurus members in an H-R diagram, which suggests that they may
be seen in scattered light. Indeed, we demonstrate in the next section
that 2MASS~J04202144+2813491 has an edge-on disk.  Because our
calculated luminosities for these objects are probably not reliable,
we do not report these estimates in Table~\ref{tab:new} and do not
plot these stars on the H-R diagram in Figure~\ref{fig:hr}.

The temperatures and luminosities of the new and previously known
members of Taurus (except for subluminous sources) are plotted on an
H-R diagram in Figure~\ref{fig:hr} with the predictions of theoretical
evolutionary models \citep{bar98,cha00}. LH~0429+17 is included among
the previously known members in Figure~\ref{fig:hr}.  A few of the new
members have distinctive positions in the H-R diagram.  FU~Tau~A and
2M~0419+2712 are the brightest known members of Taurus near their
spectral types. The overluminous nature of the former was discussed by
\citet{luh09fu}. The high luminosity of 2M~0419+2712 is consistent
with the very early evolutionary stage that is implied by the shape of
its spectral energy distribution (\S~\ref{sec:proto}).  Meanwhile, the
temperature and luminosity of the coolest known member of Taurus,
2M~0437+2331, correspond to an age of 100~Myr according to the
evolutionary models, which is much older than expected for a member of
Taurus. Stars with edge-on disks and nonmembers can appear to have
very low luminosities on the H-R diagram. However, a disk does not
appear to be present (\S~\ref{sec:ldwarf}) and the strengths of the
gravity-sensitive lines of this object provide strong evidence of
youth.  Instead, the old isochronal age for this object is likely a
reflection of errors in the adopted temperature scale and evolutionary
models.  This conclusion is based on the cluster sequence for
Chamaeleon~I, which falls along older model isochrones at spectral
types later than M8
\citep{luh08cha1}. In fact, the position of the coolest known member of 
Chamaeleon~I on the H-R diagram is very similar to that of
2M~0437+2331. Thus, the luminosity of the latter is not anomalous when
compared to other young late-type objects.  If we estimate the mass of
2M~0437+2331 based on its luminosity as done by \citet{luh08cha1} for
the coolest member of Chamaeleon~I, then we arrive at a value of
4-7~$M_{\rm Jup}$ for an assumed age range of 1-3~Myr.

\subsection{Disks}
\label{sec:disks}

We can use the extensive mid-IR images of Taurus from {\it Spitzer} to
determine whether the new Taurus members have circumstellar disks.
Photometry from IRAC (3.6-8.0~\micron) and MIPS (24~\micron) for the
new members is presented by \citet{luh09tau}, who measured photometry
for all known members of Taurus that appear within {\it Spitzer}
images of the region. IRAC and MIPS data for the nonmembers in our
spectroscopic sample are provided in Tables~\ref{tab:nonirac} and
\ref{tab:nonmips}, respectively. Using the {\it Spitzer} photometry, 
\citet{luh09tau} have classified each member of Taurus 
as class~I, II, or III, which follows the standard classification
scheme for the spectral energy distributions (SEDs) of young stars
\citep{lada87}.  The 41 new members in our spectroscopic sample were
identified as candidates through data from {\it Spitzer} (24), {\it
XMM-Newton} (16), and a companion survey (1). Although the {\it
Spitzer} members were selected based on red mid-IR colors, one of
these sources (2M~0437+2331) does not show evidence of a disk upon
closer examination of its colors later in this section.  All of the
X-ray members have been observed by IRAC and MIPS, but these data are
not yet publicly available for one object, XEST~06-006.  XEST~11-078
and XEST~26-062 are class~I and class~II, respectively, while the
remaining 13 X-ray sources are class~III. The candidate companion
2MASS~J04414565+2301580 is class~III as well.  In the remainder of
this section, we discuss in detail the evidence of disks for a few of
the new members that are particularly notable.

\subsubsection{L-type Brown Dwarf: No Disk}
\label{sec:ldwarf}

\begin{figure}
\plotone{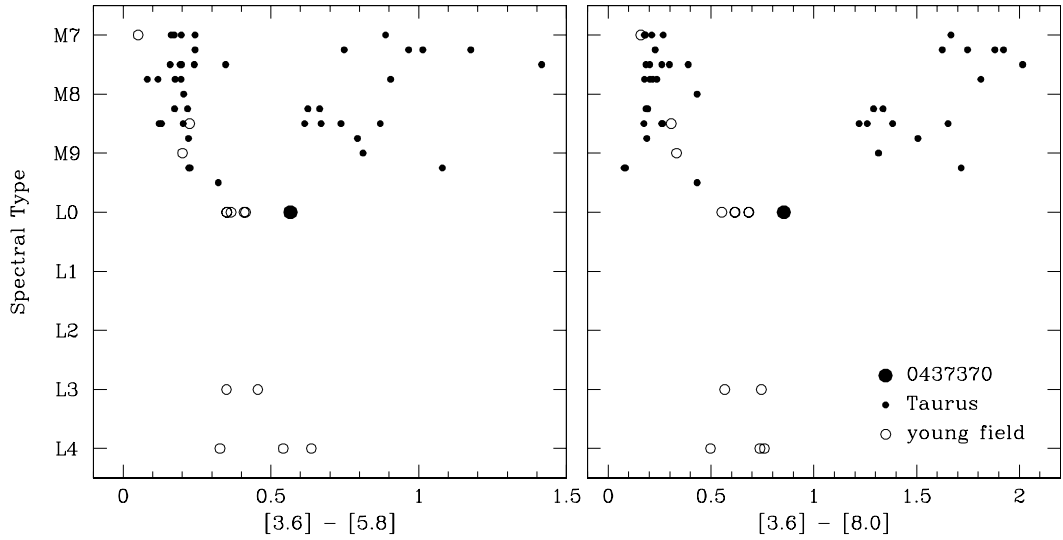}
\caption{
Spectral types versus mid-IR colors for late-type members of Taurus 
({\it filled circles}, $\tau\sim1$~Myr) and young field dwarfs 
({\it open circles}, $\tau\lesssim100$~Myr).
The blue sequence in each color represents stellar photospheres 
while the redder objects are likely to have circumstellar disks. 
The new L0 member of Taurus ({\it large filled circle}) does not exhibit
significant color excesses relative to young L-type members of the field
\citep{kir06,kir08,cru09}.
}
\label{fig:color}
\end{figure}

2M~0437+2331 is the coolest known member of Taurus and may be one of
the least massive known brown dwarfs.  It was selected for
spectroscopy based on red IRAC colors that suggested the presence of a
disk (it was not detected by MIPS). To reliably determine whether
2M~0437+2331 does indeed have a disk, we must compare its colors to
those of stellar photospheres.  For these comparison sources, we
select all known late-type members of Taurus and young late-M and L
dwarfs in the solar neighborhood
\citep{kir06,cru09} that have been observed by IRAC. 
The IRAC data for the Taurus members are taken from \citet{luh09tau}.
We have measured photometry for young field dwarfs observed in {\it
Spitzer} programs 284 (K. Cruz) and 30540 (J. Houck) with the same
methods that were employed for the Taurus data. These data are
presented in Table~\ref{tab:ldwarfs}. We show $[3.6]-[5.8]$ and
$[3.6]-[8.0]$ versus spectral type for 2M~0437+2331, other Taurus
members, and the young field dwarfs in Figure~\ref{fig:color}. These
data form two distinct groups, a narrow sequence of stellar
photospheres and a broader distribution of redder sources that have
disks.  2M~0437+2331 is only slightly redder than the photospheric
sequences ($\sim0.2$~mag) and does not exhibit color excesses as large
as those found among the Taurus members that have disks
($\sim0.5-1.5$~mag).  Thus, we do not find significant evidence of a
disk for this object. As demonstrated in Figure~\ref{fig:color}, the
photospheric IRAC colors become redder from M to L types among young
objects, which explains why 2M~0437+2331 was identified as a possible
disk-bearing object through our IRAC color criteria.  Because young
stars and brown dwarfs that lack IRAC excesses rarely exhibit evidence
of accretion in H$\alpha$
\citep{moh05,muz03,muz05,har05,gui07,luh06tau2,luh08cha1}, 
the H$\alpha$ emission in the spectrum of 2M~0437+2331 probably arises from 
the stellar chromosphere rather than accretion.

\subsubsection{Candidate Class~I Brown Dwarf}
\label{sec:proto}

\begin{figure}
\epsscale{0.65}
\plotone{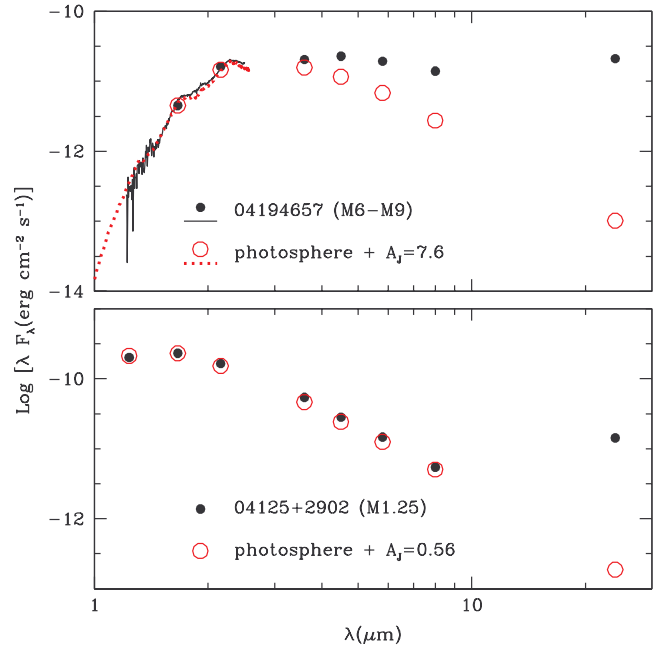}
\caption{
Two notable SEDs among the new members of Taurus ({\it filled circles and
solid line}) compared to SEDs of young stellar photospheres with similar 
spectral types ({\it open circles and dotted line}). 
2MASS~J04194657+2712552 has the highest extinction of any known late-type 
member of Taurus and exhibits a flat mid-IR SED, suggesting that it may
have a protostellar envelope. 
IRAS~04125+2902 shows excess emission at 24~\micron\ but not at
$\lambda\leq8$~\micron, which is a signature of a disk with an inner hole. 
Each photospheric SED has been reddened by the extinction of the 
Taurus source and scaled to its $H$-band flux.
}
\label{fig:sed}
\end{figure}

The IR spectrum of 2M~0419+2712 has strong H$_2$O absorption and is
highly reddened ($A_J=7.6$), making it the most heavily obscured
late-type member of Taurus found to date. To further investigate the
nature of this object, we have constructed its SED in
Figure~\ref{fig:sed} with photometry from 2MASS, IRAC, and MIPS and
our near-IR spectrum, which was flux-calibrated with the 2MASS
data. The Taurus member KPNO~5 is similar to 2M~0419+2712 in spectral
type and does not have mid-IR excess emission
\citep{bri02,har05,luh06tau2}. Therefore, we include the SED of KPNO~5 in 
Figure~\ref{fig:sed} as an estimate of the stellar photosphere of 
2M~0419+2712. The SED of KPNO~5 consists of data from 2MASS, SpeX
\citep{mue07}, and {\it Spitzer} \citep{luh09tau} and has been reddened by 
$A_J=7.3$ to match the reddening of 2M~0419+2712 (KPNO~5 has 
$A_J\sim0.3$). We use the reddening laws from \citet{rl85} and \citet{fla07}.
As shown in Figure~\ref{fig:sed}, 2M~0419+2712 exhibits significant
excess emission relative to KPNO~5 beyond 3~\micron. The SED of 2M~0419+2712
is roughly flat from 2-24~\micron, which is suggestive of an
evolved class~I source (star+disk+envelope) or a very young class~II object
(star+disk). Thus, 2M~0419+2712 may offer a rare opportunity for studying
a brown dwarf in the protostellar phase. A definitive classification of 
the evolutionary stage of this object will require additional observations, 
such as mid-IR spectroscopy \citep{fur08}.

\subsubsection{Candidate Transitional Disk}

Most of the candidate members that we identified with {\it Spitzer}
photometry have red colors in all available bands beyond
3~\micron. However, one of the confirmed members from that sample,
IRAS~04125+2902, is red only at 24~\micron.  To illustrate the
distinctive colors of this star, we compare its SED to the average SED
of early-M diskless stars in Chamaeleon~I and Taurus in
Figure~\ref{fig:sed} \citep{luh08cha1,luh09tau}.  The data for
IRAS~04125+2902 agree well the photospheric SED in the 2MASS and IRAC
bands, but they show significant excess emission at 24~\micron. An SED
of this kind indicates the presence of a disk with an inner hole,
which is known as a transitional disk \citep{cal02,cal05,dal05}.
Mid-IR spectroscopy and millimeter imaging of this disk are needed for
a detailed characterization of the inner hole
\citep{esp07a,esp07b,fur07,bro08,dut08,hug07,hug09}.

\subsubsection{Edge-on Disk}

\begin{figure}
\epsscale{0.55}
\plotone{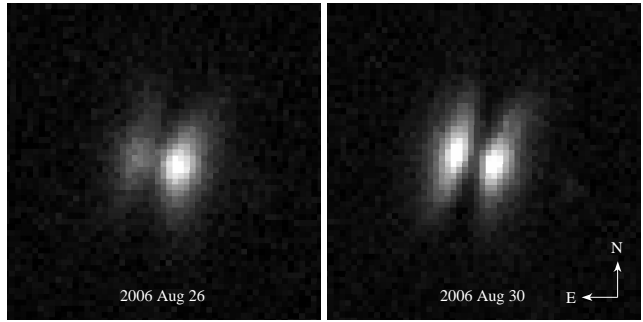}
\caption{
$I$-band CFHT images of the new Taurus member 2MASS~J04202144+2813491
($10\arcsec\times10\arcsec$). 
These data exhibit bipolar extended emission separated by a dark lane,
indicating the presence of an edge-on circumstellar disk that is occulting
the star. The brightness of the eastern lobe of scattered light varied
significantly between these two dates.
Point sources in the surrounding areas of these images
exhibit FWHM$=0\farcs7$ ({\it left}) and $0\farcs5$ ({\it right}). 
}
\label{fig:edge}
\end{figure}

As noted in the previous section, a few of the new Taurus members are
much fainter at optical and near-IR wavelengths than most other
members with similar spectral types, indicating that they may have
edge-on disks. We have investigated this possibility by inspecting
images of these stars that are available in the data archives of
various observatories and wide-field surveys.  In optical images from
MegaPrime/MegaCam at the Canada-France-Hawaii Telescope (CFHT),
2MASS~J04202144+2813491 does indeed exhibit a clear signature of an
edge-on disk in the form of two lobes of extended emission that are
separated by a dark lane, as shown in Figure~\ref{fig:edge}. The
radius of the disk is $2\farcs5$, corresponding to 350~AU at the
distance of Taurus.  After comparing CFHT images that were collected
on different nights, we find that the scattered light from the disk
experienced significant variability over a period of a few days (see
Figure~\ref{fig:edge}). Similar variability has been detected in the
scattered light from HH~30 \citep{bur96,wat07}.
The discovery of 2MASS~J04202144+2813491 illustrates the utility of
{\it Spitzer} imaging for finding edge-on disks, which can be
overlooked through other types of surveys for young stars
\citep{luh08cha2}.

\subsection{Companions}

A few of the new Taurus members have candidate companions at wide
separations.  2MASS~J04414565+2301580 is located $12\arcsec$ from the
known member 2MASS~J04414489+2301513 \citep[M8.5][]{luh06tau1} and was
identified as a possible companion by \citet{kra07}. We have confirmed
its youth and membership in Taurus through spectroscopy.  Our spectral
classification of M4-M5 for 2MASS~J04414565+2301580 is consistent with
that expected for a coeval companion to 2MASS~J04414489+2301513 based
on the evolutionary models of \citet{bar98}.  While performing
spectroscopy on FU~Tau, we noticed a fainter object at a separation of
$5.7\arcsec$, which we confirmed spectroscopically as a young brown
dwarf. We discussed this pair in detail in a separate study
\citep{luh09fu}.

By inspecting images of the remaining new members, we have identified
two additional candidate companions.  The new member IRAS~04125+2902
is $4\arcsec$ from a source that is $\sim2$~mag fainter and is
detected by DSS, 2MASS, and IRAC. This candidate,
2MASS~J04154269+2909558, does not exhibit mid-IR excess emission in
the IRAC data, indicating that it is a class~III source if it is a
member of Taurus. The second candidate companion,
2MASS~J04355949+2238291, is $10\arcsec$ from XEST~09-042. Although it
has red mid-IR colors, it is probably a galaxy rather than a young
star based on the shape of its SED.

\section{Initial Mass Function}
\label{sec:imf}

\begin{figure}
\epsscale{0.7}
\plotone{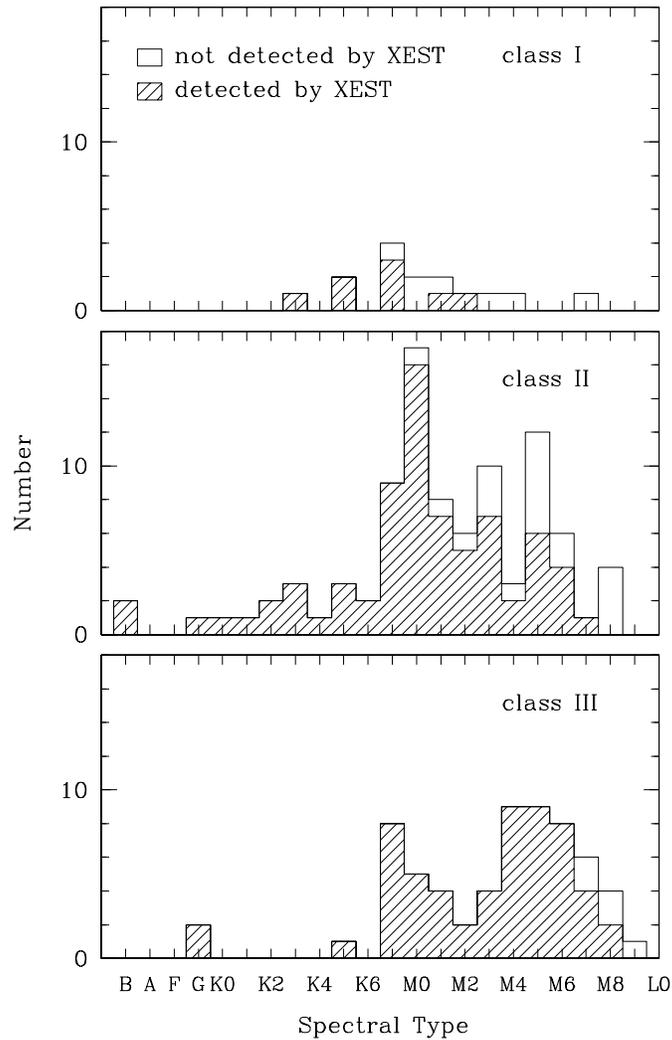}
\caption{
Distributions of spectral types for known members of Taurus that are within 
the {\it XEST} fields as a function of SED class. The stars that were 
detected by {\it XMM} are indicated ({\it shaded histograms}).
}
\label{fig:histox}
\end{figure}

We wish to use our updated census of Taurus to estimate the IMF in
this region.  The completeness of the current census is a function of
several parameters, including stellar mass, location, SED class, and
extinction.  Therefore, we must carefully define the sample of members
for inclusion in the IMF so that it is representative of the stellar
population in Taurus.  We begin by considering members of Taurus that
are located within the areas observed by XEST, which are shown in
Figure~\ref{fig:map}. We select the XEST fields because they encompass
a large number of members and are almost entirely covered by deep
imaging at optical and IR wavelengths as well as in X-rays
\citep{bri02,luh04tau,luh06tau1,luh06tau2,gui06}. 
To evaluate the completeness of the census of Taurus within these fields, 
we first examine the completeness of XEST.
In Figure~\ref{fig:histox}, we plot the distributions of spectral types 
for all known members of Taurus within the XEST images and for
the members detected in those data. Separate histograms are shown for 
classes~I, II, and III, where the SED classifications are from 
\citet{luh09tau}. We omit secondaries that are unresolved by XEST.
An additional 16 members are absent from Figure~\ref{fig:histox}
because they lack accurate spectral types, five of which were detected
by XEST.  All of these unclassified stars have class~I SEDs.  Only 15
of the 31 class~I sources within the XEST fields have spectral
classifications and thus are present in Figure~\ref{fig:histox}.
According to Figure~\ref{fig:histox}, the completeness of XEST
decreases for lower stellar masses and earlier SED classes, which is a
reflection of the fact that X-ray emission is correlated with both of
these properties \citep{tel07,pri08}. XEST is nearly 100\% complete
for class~III members earlier than M6 ($M\sim0.1$~$M_\odot$).
\citet{gro07} reported that XEST detected 8 of 16 members later than M6 that 
are within those images. Using our updated census of Taurus and the spectral 
types that we have measured in this work and in previous studies, we find that 
the XEST detection fraction is 7/17 for types later than M6.

Optical and IR surveys for new members of Taurus have complemented
XEST in terms of completeness. Images from {\it Spitzer} have covered
$\sim$92\% of the XEST fields \citep{luh09tau}. The current census for
that portion of the XEST fields is nearly complete for class~I and II
stars and for class~II brown dwarfs down to $\sim0.02$~$M_\odot$
according to {\it Spitzer} surveys for new members \citep[][this
work]{luh06tau2}. The completeness for class~I brown dwarfs is unknown
because of contamination of the {\it Spitzer} images by red galaxies
\citep{luh06tau2}. The remaining $\sim8$\% of the XEST fields that
lacks {\it Spitzer} images does not encompass any known members or
dark clouds, and thus probably does not contain a significant number
of undiscovered class~I and II sources. Meanwhile, the XEST fields
have been fully covered at optical and near-IR wavelengths by either
deep wide-field imaging
\citep{bri02,luh04tau,gui06} or all-sky catalogs \citep{luh06tau1}, 
which have produced a high level of completeness 
for classes II and III between 0.1-0.02~$M_\odot$ within an extinction range
that encompasses most members \citep[$A_V<4$;][]{luh04tau,luh06tau1,gui06}.
Thus, the X-ray, optical, and IR studies have produced a
census of Taurus members within the areas covered by XEST that should be 
complete down to masses of $\sim0.02$~$M_\odot$ for classes II and III.

\begin{figure}
\epsscale{0.6}
\plotone{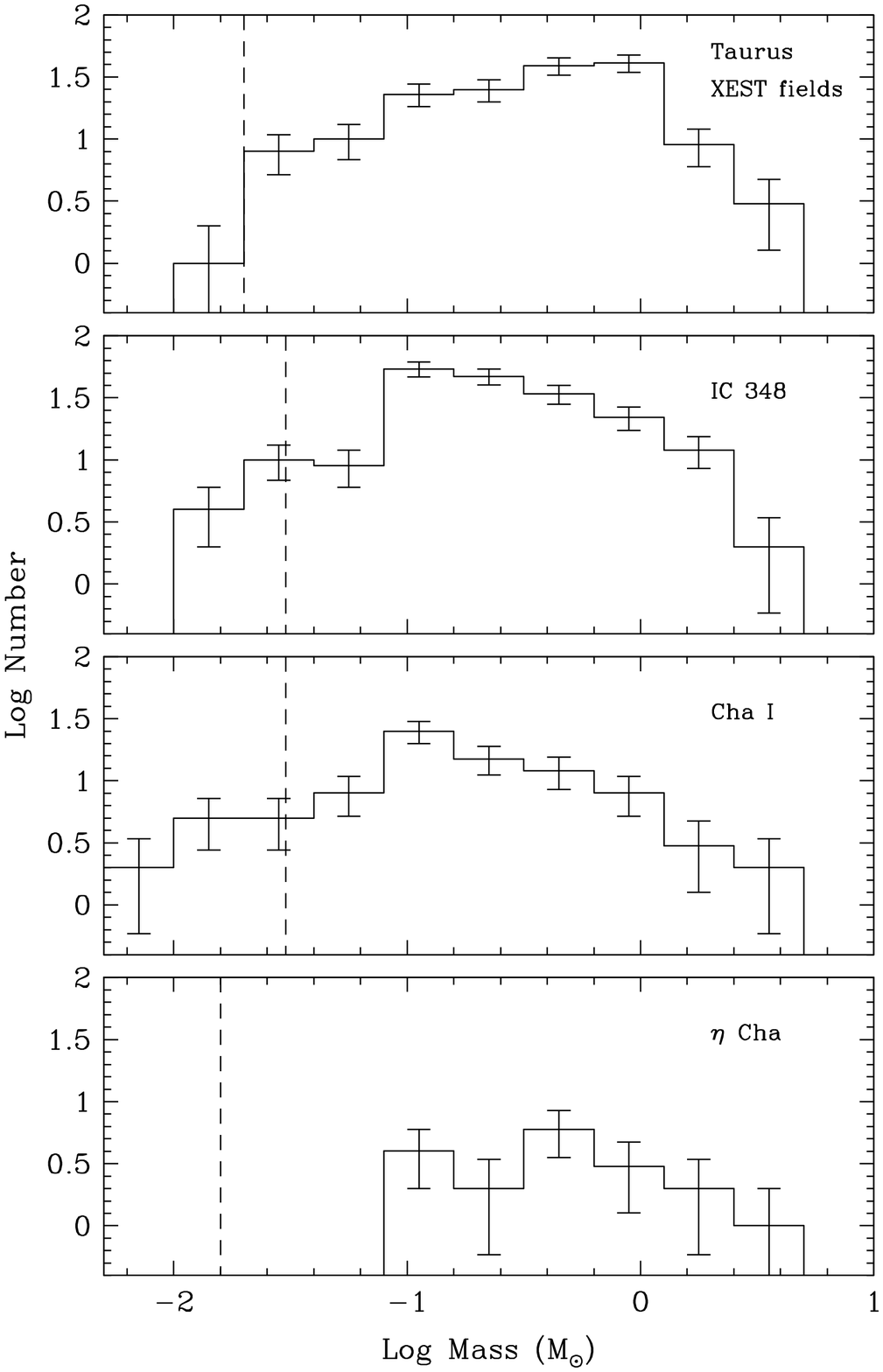}
\caption{
IMFs for Taurus, IC~348 \citep{luh03ic}, and Chamaeleon~I \citep{luh07cha}.
The IMF for Taurus is derived from members within the XEST fields 
(Figure~\ref{fig:map}). These IMFs contain 157, 194, and 85 sources,
respectively. The Taurus IMF differs significantly from the mass functions
in the other two regions (\S~\ref{sec:imf}).
We also include the IMF of the $\eta$~Cha association
\citep{mam99,law02,lyo04,sz04,luh04eta1}.
Although \citet{lyo04} reported that $\eta$~Cha exhibits a deficit of
low-mass stars and brown dwarfs relative to other clusters, its mass function
is statistically consistent with the IMFs of IC~348
and Chamaeleon~I \citep[][\S~\ref{sec:imf}]{luh04eta2}.
The completeness limits of these samples are indicated ({\it dashed lines}). 
In the units of this diagram, the Salpeter slope is 1.35.
}
\label{fig:imf}
\end{figure}

\begin{figure}
\epsscale{0.6}
\plotone{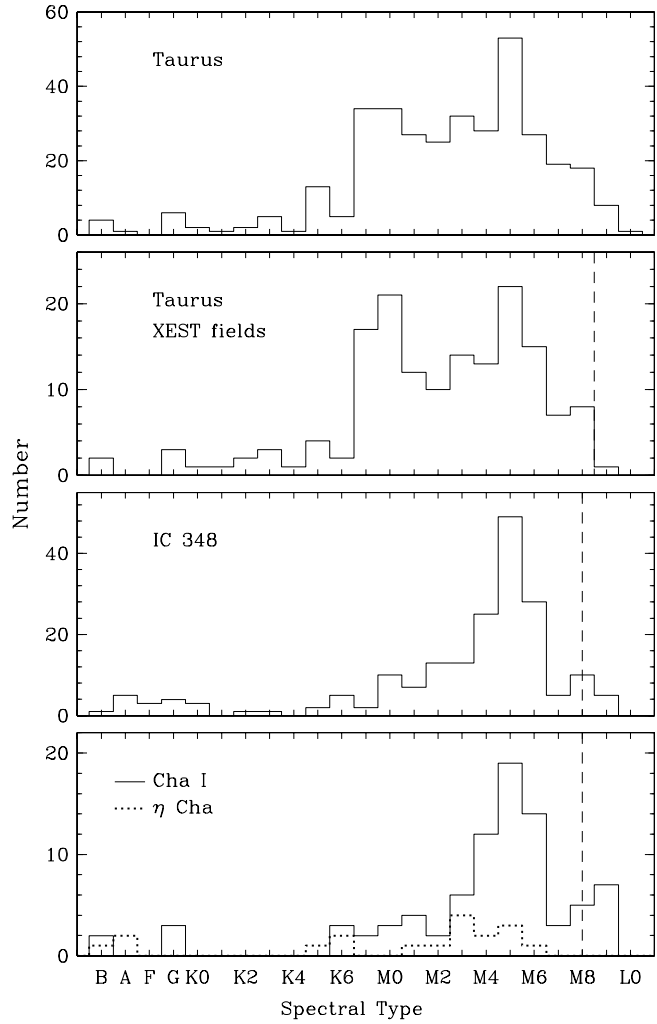}
\caption{
Distributions of spectral types for all known members of Taurus and for the
IMFs of Taurus, IC~348, Chamaeleon~I, and $\eta$~Cha in Figure~\ref{fig:imf}.
The completeness limits of the IMF samples in Taurus, IC~348, and Chamaeleon~I 
are indicated ({\it dashed lines}). The limit for $\eta$~Cha is near M9.
}
\label{fig:histo}
\end{figure}

As done for the H-R diagram in \S~\ref{sec:hr}, we treat multiple
systems that are unresolved by 2MASS as single sources when
constructing the IMF for the XEST fields. We exclude class~I sources,
stars that lack accurate spectral classifications (most of which are
class~I), and objects that are probably seen in scattered light based
on their unusually low luminosity estimates. The resulting mass
function consists of primaries and widely separated secondaries, as in
our previous measurements of IMFs in Taurus and other star-forming
regions \citep{luh03ic,luh04tau,luh07cha}.  We have estimated the
masses for the Taurus members in the IMF sample from their positions
on the H-R diagram in Figure~\ref{fig:hr} by using the theoretical
evolutionary models of \citet{bar98} and \citet{cha00} for
$M/M_\odot\leq1$ and the models of \citet{pal99} for $M/M_\odot>1$.
The IMF for the XEST fields is presented in Figure~\ref{fig:imf} in
logarithmic units where the Salpeter slope is $\alpha=1.35$.  It
exhibits a maximum near 0.8~$M_\odot$ and declines steadily to lower
masses ($\alpha=-0.44$), and thus closely resembles the IMFs that we
have previously reported for Taurus \citep{bri02,luh04tau}. In
comparison, the mass functions of other nearby star-forming regions
peak at 0.1-0.2~$M_\odot$
\citep{hil97,hc00,mue02,mue03,luh03ic,luh07cha}. 
This variation in the IMF is illustrated in Figure~\ref{fig:imf},
where we include mass functions for IC~348 and Chamaeleon~I that were
derived in the same manner as our measurement for Taurus
\citep{luh03ic,luh07cha}.  The surplus of stars near 0.8~$M_\odot$
(K7-M1) in Taurus relative to those two clusters is also apparent in
the distributions of spectral types for the IMF samples in
Figure~\ref{fig:histo}.  Based on a two-sided Kolmogorov-Smirnov test,
the probability that the sample for Taurus was drawn from the same
mass distribution as either IC~348 or Chamaeleon~I is
$\sim0.04$\%. Possible explanations for the distinctive shape of the
IMF in Taurus have been discussed in previous studies and generally
involve a higher average Jeans mass in this region
\citep{bri02,luh03ic,luh04tau,goo04,lada08}.

Finally, we comment briefly on another young population in which an
unusual IMF has been reported. By constructing an IMF from all known
members of the $\eta$~Cha association
\citep{mam99,law02,lyo04,sz04,luh04eta1},
\citet{lyo04,lyo06} found that $\eta$~Cha exhibits a deficit of 20--29 
low-mass stars and brown dwarfs (0.025--0.15~$M_\odot$) relative to the
solar neighborhood and other young clusters. \citet{mor07} attempted to 
provide a theoretical explanation for the apparently unusual IMF in this
association. However, through a survey of $\eta$~Cha that was complete to 
0.015~$M_\odot$\footnote{\citet{lyo06} and \citet{mor07} incorrectly quoted
a completeness limit of 0.025~$M_\odot$ for the surveys by \citet{luh04eta1}
and \citet{luh04eta2}, which employed 
photometry from DENIS and 2MASS. \citet{lyo06} suggested that 
the completeness limit from those surveys were determined by their shallowest
data, which consisted of the optical photometry from DENIS. However, any
objects above 0.015~$M_\odot$ that were absent from the optical data 
would have appeared in the IR diagrams used by \citet{luh04eta1} and 
\citet{luh04eta2} for selecting candidate members. Thus, the completeness 
limit was 0.015~$M_\odot$.},
\citet{luh04eta1} and \citet{luh04eta2} concluded that a significant 
paucity of low-mass objects is not present.  To further explore this
issue, we have constructed an IMF and a distribution of spectral types
for all known members of $\eta$~Cha, which are included in
Figures~\ref{fig:imf} and \ref{fig:histo}, respectively. A two-sided
Kolmogorov-Smirnov test indicates a probability of $\sim10$\% that the
members of $\eta$~Cha are drawn from the same mass distribution as
either IC~348 or Chamaeleon~I, which does not represent a significant
difference. Thus, the IMF in $\eta$~Cha is consistent with the mass
functions in IC~348 and Chamaeleon~I.

\section{Conclusions}

We have presented a survey for new members of the Taurus star-forming
region in which we obtained spectra of candidate members appearing in
images from the {\it Spitzer Space Telescope} (46~deg$^{2}$) and the
{\it XMM-Newton Observatory} (5~deg$^{2}$).  Using the mid-IR data
from {\it Spitzer}, we identified 44 sources that could be young stars
with disks, 24 of which were confirmed as members by our
spectroscopy. We also performed spectroscopy on 51 candidates detected
in X-rays by the XEST program \citep{gud07,sce07}, demonstrating the
youth and membership of 16 sources. Ten of these new X-ray members
were independently confirmed through spectroscopy by \citet{sce08}.
In addition, to the sources from {\it Spitzer} and {\it XMM-Newton},
we observed four candidate companions to known members of Taurus that
were found by \citet{kra07} through analysis of 2MASS data, one of
which we have classified as a young star.

Our survey has uncovered several rare types of sources that are
valuable for studies of various aspects of star and planet
formation. They consist of a wide binary brown dwarf that is forming
in isolation \citep{luh09fu}, the first known L-type member of Taurus
($M\sim4$-7~$M_{\rm Jup}$), a highly reddened brown dwarf that may be
in the class~I stage (star+disk+envelope), a disk that appears to have
an inner hole (i.e., transitional disk), and a large edge-on disk
($r=2\farcs5=350$~AU).  The companion identified by \citet{kra07} also
may comprise the primary in wide, low-mass binary system (M4.5+M8.5,
$a=12\arcsec=1700$~AU).  Meanwhile, the {\it Spitzer} and {\it
XMM-Newton} data in conjunction with previous optical and near-IR
surveys provide relatively well-defined completeness limits for the
current census of Taurus, enabling a better characterization of the
stellar population.  For instance, we have estimated the IMF within
the fields observed by XEST, arriving at a distribution that reaches a
maximum near 0.8~$M_\odot$, which agrees with our previous
measurements for Taurus. Thus, the IMF in Taurus continues to appear
anomalous compared to other nearby star-forming clusters, which peak
at 0.1-0.2~$M_\odot$.  The disk fraction in the XEST fields and the
spatial distribution of the SED classes are investigated by
\citet{luh09tau}.

The completeness of the census of Taurus remains poorly determined
among class~I sources at low masses and class~III sources outside of
the XEST fields, which focused on the denser stellar
aggregates. Future surveys can address these shortcomings through
spectroscopy of red, faint sources detected by {\it Spitzer} and
measurements of variability and proper motions with wide-field,
multi-epoch imaging (e.g., Panoramic Survey Telescope and Rapid
Response System, Large Synoptic Survey Telescope).

\acknowledgements

K.~L. was supported by grant AST-0544588 from the National Science
Foundation. This work makes use of data from the {\it Spitzer Space
Telescope} and 2MASS.  {\it Spitzer} is operated by the Jet Propulsion
Laboratory, California Institute of Technology under a contract with
NASA.  2MASS is a joint project of the University of Massachusetts and
the Infrared Processing and Analysis Center/California Institute of
Technology, funded by NASA and the NSF.  Support for K.~C. was
provided by NASA through the {\it Spitzer Space Telescope} Fellowship
Program through a contract issued by JPL/Caltech.  {\it XMM-Newton} is
an ESA science mission with instruments and contributions directly
funded by ESA Member States and NASA.  The HET is a joint project of
the University of Texas at Austin, the Pennsylvania State University,
Stanford University, Ludwig-Maximillians-Universit\"at M\"unchen, and
Georg-August-Universit\"at G\"ottingen. The HET is named in honor of
its principal benefactors, William P. Hobby and Robert E. Eberly. The
Marcario Low-Resolution Spectrograph at HET is named for Mike Marcario
of High Lonesome Optics, who fabricated several optics for the
instrument but died before its completion; it is a joint project of
the HET partnership and the Instituto de Astronom\'{\i}a de la
Universidad Nacional Aut\'onoma de M\'exico.  MegaPrime/MegaCam is a
joint project of CFHT and CEA/DAPNIA. CFHT is operated by the National
Research Council of Canada, the Institute National des Sciences de
l'Univers of the Centre National de la Recherche Scientifique of
France, and the University of Hawaii. Gemini Observatory is operated
by AURA under a cooperative agreement with the NSF on behalf of the
Gemini partnership: the NSF (United States), the Particle Physics and
Astronomy Research Council (United Kingdom), the National Research
Council (Canada), CONICYT (Chile), the Australian Research Council
(Australia), CNPq (Brazil) and CONICET (Argentina). This research has
made use of data obtained from the SuperCOSMOS Science Archive,
prepared and hosted by the Wide Field Astronomy Unit, Institute for
Astronomy, University of Edinburgh, which is funded by the UK Science
and Technology Facilities Council.

\appendix

\section{Adopted Census and Kinematics of Taurus} 
\label{sec:app}

\begin{figure}
\epsscale{1}
\plotone{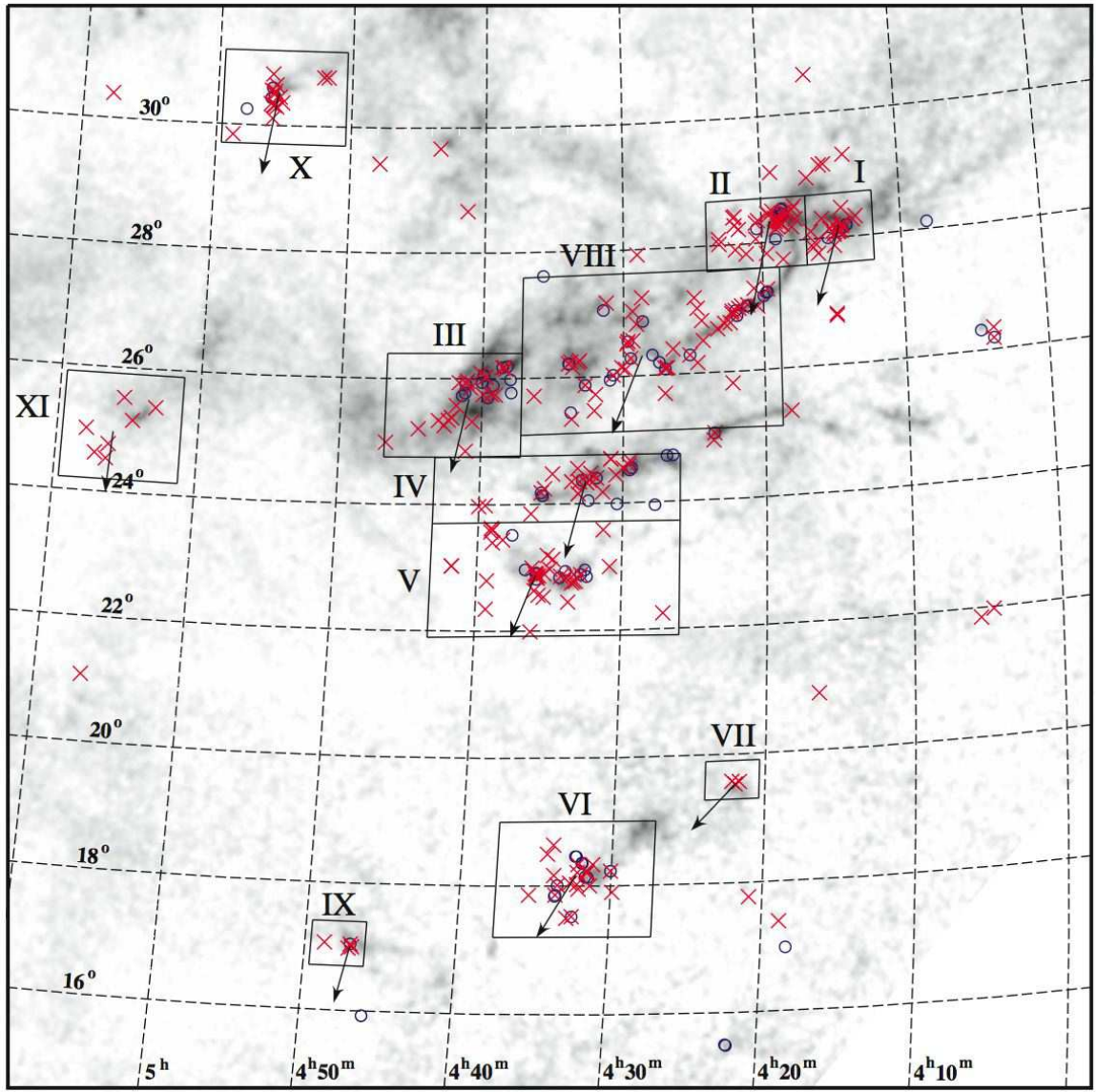}
\caption{
Spatial distribution of all known members of the Taurus star-forming region,
which are labeled according to the presence or absence of proper motion
measurements ({\it crosses and circles}).
The median proper motions of members within 11 groups
({\it rectangles}) are presented in Table~\ref{tab:pm}.
The corresponding motions over a period of 0.2~Myr are indicated ({\it arrows}).
The dark clouds in Taurus are displayed with a map of extinction 
\citep[{\it grayscale},][]{dob05}.
}
\label{fig:pm}
\end{figure}

Our analysis of the IMF in Taurus required a census of all known
members of the region. While constructing this census, we have made
use of proper motion measurements when they are available.  For a star
whose membership is uncertain, we have compared its proper motion to
that of the nearest group of known members. In Table~\ref{tab:pm}, we
present the median proper motions of 11 groups in Taurus based on
stars from our adopted census and proper motions measurements from a
variety of sources
\citep{harr99,hog00,ham01,han04,zac04a,zac04b,duc05,loi07,van07,ros08,tor07,tor09}.
Our adopted boundaries for these groups and the median motions are
shown on a map of Taurus in Figure~\ref{fig:pm}.

We describe our adopted census of known members of Taurus in terms of
modifications to the compilation of members presented by \citet{ken08}
in a review of this star-forming region. We begin by identifying the
sources from that list that we have excluded as members. The proper
motions of HBC~351 ($\mu_{\alpha}, \mu_{\delta}= +18.5\pm2.0$,
$-49.2\pm2.0$~mas~yr$^{-1}$), HBC~352 ($+6\pm1$,
$-9\pm1$~mas~yr$^{-1}$), HBC~353 ($+7\pm8$, $-12\pm2$~mas~yr$^{-1}$),
and HBC~356/357 ($-1\pm5$, $-9\pm5$~mas~yr$^{-1}$) from
\citet{zac04a,zac04b} differ significantly from those of the nearest
Taurus groups in Table~\ref{tab:pm}.  The available proper motion
measurements for HBC~354 and HBC~355, which are separated by
$6\arcsec$, also are inconsistent with membership in Taurus
\citep{zac04a,duc05,ros08}. We provided some of the new members
reported in this work to \citet{ken08} for inclusion in their list of
members, including 2MASS~J04345973+2807017 and
2MASS~J04124858+2749563. However, upon closer examination of these two
stars, we have concluded that they are probably not members of Taurus
(\S~\ref{sec:comm}). Our spectroscopy indicates that IRAS~04428+2403
is a galaxy. No evidence of membership has been presented for
V410~Anon~20 and V410~Anon~24.  They are fainter than other Taurus
members near their spectral types, suggesting that they are background
stars. CIDA-13 and St34 have been classified as probable foreground
stars \citep{muz03,har05st}.

In addition to the members compiled by \citet{ken08}, we have included
in our census the new members presented in Table~\ref{tab:new} as well
as 2MASS J04162725+2053091, 2MASS J04270739+2215037, 2MASS
J04344544+2308027, 2MASS J04381630+2326402, 2MASS J04385859+2336351,
2MASS J04385871+2323595, 2MASS J04390163+2336029, 2MASS
J04390637+2334179 \citep{sle06}, 2MASS J04295422+1754041, 2MASS
J04263055+2443558 \citep{luh06tau1}, IRAS~04325+2402C \citep{har99},
IRAS~04111+2800G \citep{pru92}, L1521F-IRS \citep{bou06},
IRAM~04191+1522 \citep{and99}, IRAS~04278+2253~B \citep{whi04}, 
LH~0429+17 \citep{harr99,rei99}, and HD~28867 \citep{wal03}.
2MASS J04333278+1800436 was identified as a possible member of Taurus
by \citet{wal03} based on its close proximity to HD~28867.
We add it to our census since it exhibits mid-IR excess emission that
indicates the presence of a disk \citep{luh09tau}.  The proper motions of
HD~30171 \citep[$+12.1\pm1.1$,$-17.7\pm1.1$~mas~yr$^{-1}$,][]{hog00} 
and MWC~480 \citep[$+5.5\pm1.1$,$-25.4\pm1.1$~mas~yr$^{-1}$,][]{hog00} are
similar to those of the nearest Taurus groups (Table~\ref{tab:pm}),
supporting their membership in Taurus. 

We have found a few other differences between our census and that of
\citet{ken08}. The coordinates that we retrieved from the 2MASS Point 
Source Catalog differ from those in \citet{ken08} by $2-22\arcsec$ for
IRAS~04016+2610, HBC~412, IRAS~04370+2559, MHO~9, CIDA-12, and
IRAS~04191+1523.  Our adopted 2MASS counterparts for these stars are
given in \citet{luh09tau}.  IRAS~04181+2655 appears in the census from
\citet{ken08}, but it is a blend of three other stars in that list,
2MASS J04210795+2702204, 2MASS J04211038+2701372, and 2MASS
J04211146+2701094, and thus does not require a separate
entry. Similarly, IRAS~04263+2426 and GV~Tau are listed separately by
\citet{ken08} even though they represent the same star. The
coordinates for IRAS~04166+2706 in \citet{ken08} apply to
IRAS~04166+2708.  We have measured coordinates of $\alpha=4^{\rm
h}19^{\rm m}42.5^{\rm s}$, $\delta=27\arcdeg13\arcmin36.7\arcsec$
(J2000) for the true counterpart to IRAS~04166+2706 using images from
{\it Spitzer}.  The misidentification for IRAS~04166+2706 in
\citet{ken08} probably originated in \citet{luh06tau1} and
\citet{luh06tau2}.
Our adopted list of Taurus members is provided in \citet{luh09tau}.

By combining our membership list with previous astrometric and kinematic
measurements for these stars, we can constrain both the radial velocities and 
three-dimensional velocities of the Taurus groups in Table~\ref{tab:pm}.
For each group, we have calculated the median radial velocity of stars
\citep{app88,bbf94,bbf00,fin84,gah99,har86,har87,her77,mal06,mar05,mat97,
mun83,rei90,sar98,wal86,wal88,whi03,zai85,zai90} and 
the median radial velocity of the dense gas based on observations of
C$^{18}$O, CS, N$_2$H$^+$, and H$^{13}$CO$^+$ \citep{lee99,oni02,tat04}.  
The gas radial velocity for group IX is from \citet{ung87}.
The gas velocities were transformed to the heliocentric frame using the
standard solar motion \citep{ker86}.
The median values of the stellar and gas radial velocities differ by
an average of $-0.5\pm$0.3 km\,s$^{-1}$ (RV$_{star}-$RV$_{gas}$)
and agree within 2.5~$\sigma$ for all groups, providing further evidence that 
the stars and gas in Taurus share similar motions \citep[e.g.,][]{her77,har86}.
Using the median positions, proper motions, and gas radial velocities 
(except for group VII, where RV$_{star}$ is adopted), 
we have calculated velocities of the Taurus groups in
the directions of the Galactic center ($U$), 
Galactic rotation ($V$), and the North Galactic Pole ($W$).
Since accurate distances are available only for a few individual stars
\citep{tor09}, we have adopted a distance of $140\pm10$~pc for each group. 
The uncertainties in each velocity component are 
$\pm$0.9-1.5 km\,s$^{-1}$ with a mean error 1.0 km\,s$^{-1}$.  
We derive a mean velocity for the Taurus complex of $U, V, W = -15.7\pm0.7,
-11.3\pm0.7, -10.1\pm0.7$~km\,s$^{-1}$, which agrees with the mean value from 
\citet{ber07}.
However, the one-dimensional velocity dispersions among the groups 
($\sim$1 km\,s$^{-1}$) are much smaller than those estimated by 
\citet[][$\sim$6 km\,s$^{-1}$]{ber07}, who included a significant population of
off-cloud sources. The groups appear to have rather
coherent motions, suggesting that many of the off-cloud sources are
unrelated to the Taurus dark clouds.

\clearpage

\LongTables

\begin{deluxetable}{llllll}
\tablewidth{0pt}
\tablecaption{Observing Log\label{tab:log}}
\tablehead{
\colhead{} &
\colhead{} &
\colhead{} &
\colhead{} &
\colhead{$\lambda$} &
\colhead{} \\
\colhead{Night} &
\colhead{Date} &
\colhead{Telescope + Instrument} &
\colhead{Disperser} &
\colhead{($\mu$m)} &
\colhead{$\lambda/\Delta \lambda$}
}
\startdata
1 & 2006 Sep 25 & HET + LRS & G3 grism & 0.63-0.91 & 1100 \\
2 & 2006 Oct 2 & HET + LRS & G3 grism & 0.63-0.91 & 1100 \\
3 & 2006 Oct 3 & HET + LRS & G3 grism & 0.63-0.91 & 1100 \\
4 & 2006 Oct 19 & HET + LRS & G3 grism & 0.63-0.91 & 1100 \\
5 & 2006 Oct 20 & HET + LRS & G3 grism & 0.63-0.91 & 1100 \\
6 & 2006 Oct 22 & HET + LRS & G3 grism & 0.63-0.91 & 1100 \\
7 & 2006 Nov 17 & HET + LRS & G3 grism & 0.63-0.91 & 1100 \\
8 & 2006 Nov 26 & HET + LRS & G3 grism & 0.63-0.91 & 1100 \\
9 & 2007 Feb 19 & IRTF + SpeX & prism & 0.8-2.5 & 100 \\
10 & 2007 Aug 13 & HET + MRS & 79 l/mm grating & 0.45-0.9 & 11000 \\
11 & 2007 Aug 28 & HET + LRS & G3 grism & 0.63-0.91 & 1100 \\
12 & 2007 Sep 15 & HET + LRS & G3 grism & 0.63-0.91 & 1100 \\
13 & 2007 Sep 26 & HET + MRS & 79 l/mm grating & 0.45-0.9 & 11000 \\
14 & 2007 Oct 3 & HET + MRS & 79 l/mm grating & 0.45-0.9 & 11000 \\
15 & 2007 Oct 10 & HET + MRS & 79 l/mm grating & 0.45-0.9 & 11000 \\
16 & 2007 Oct 20 & HET + MRS & 79 l/mm grating & 0.45-0.9 & 11000 \\
17 & 2007 Oct 22 & HET + LRS & G3 grism & 0.63-0.91 & 1100 \\
18 & 2007 Oct 23 & HET + LRS & G3 grism & 0.63-0.91 & 1100 \\
19 & 2007 Oct 24 & HET + LRS & G3 grism & 0.63-0.91 & 1100 \\
20 & 2007 Oct 25 & HET + LRS & G3 grism & 0.63-0.91 & 1100 \\
21 & 2007 Oct 26 & HET + LRS & G3 grism & 0.63-0.91 & 1100 \\
22 & 2007 Nov 1 & HET + LRS & G3 grism & 0.63-0.91 & 1100 \\
23 & 2007 Nov 3 & HET + LRS & G3 grism & 0.63-0.91 & 1100 \\
24 & 2007 Nov 3 & HET + MRS & 79 l/mm grating & 0.45-0.9 & 11000 \\
25 & 2007 Nov 4 & HET + LRS & G3 grism & 0.63-0.91 & 1100 \\
26 & 2007 Nov 13 & HET + LRS & G3 grism & 0.63-0.91 & 1100 \\
27 & 2007 Nov 15 & HET + LRS & G3 grism & 0.63-0.91 & 1100 \\
28 & 2007 Nov 28 & HET + LRS & G3 grism & 0.63-0.91 & 1100 \\
29 & 2007 Dec 3 & HET + LRS & G3 grism & 0.63-0.91 & 1100 \\
30 & 2007 Dec 3 & IRTF + SpeX & prism & 0.8-2.5 & 100 \\
31 & 2007 Dec 14 & HET + LRS & G1 grism & 0.57-1.03 & 750 \\
32 & 2007 Dec 17 & Magellan~II + LDSS-3 & VPH all grism & 0.58-1.1 & 750 \\
33 & 2007 Dec 18 & Magellan~II + LDSS-3 & VPH red grism & 0.6-1 & 1400 \\
34 & 2007 Dec 20 & HET + MRS & 79 l/mm grating & 0.45-0.9 & 11000 \\
35 & 2007 Dec 23 & HET + LRS & G1 grism & 0.57-1.03 & 750 \\
36 & 2008 Jan 6 & HET + LRS & G1 grism & 0.57-1.03 & 750 \\
37 & 2008 Jan 10 & HET + LRS & G1 grism & 0.57-1.03 & 750 \\
38 & 2008 Jan 13 & HET + MRS & 79 l/mm grating & 0.45-0.9 & 11000 \\
39 & 2008 Jan 13 & HET + LRS & G1 grism & 0.57-1.03 & 750 \\
40 & 2008 Jan 15 & HET + MRS & 79 l/mm grating & 0.45-0.9 & 11000 \\
41 & 2008 Feb 17 & HET + MRS & 79 l/mm grating & 0.45-0.9 & 11000 \\
42 & 2008 Sep 1 & Gemini + GMOS & 400 l/mm grating & 0.56-1 & 1500 \\
43 & 2008 Sep 13 & HET + LRS & G3 grism & 0.63-0.91 & 1100 \\
44 & 2008 Sep 30 & Gemini + NIRI & f/6 K grism & 1.9-2.5 & 500 \\
45 & 2008 Oct 2 & HET + LRS & G3 grism & 0.63-0.91 & 1100 \\
46 & 2008 Oct 10 & Gemini + NIRI & f/6 K grism & 1.9-2.5 & 500 \\
47 & 2008 Oct 23 & HET + LRS & G3 grism & 0.63-0.91 & 1100 \\
48 & 2008 Nov 2 & IRTF + SpeX & prism & 0.8-2.5 & 100 \\
49 & 2009 Feb 12 & IRTF + SpeX & prism & 0.8-2.5 & 100 \\
50 & 2009 Feb 14 & HET + LRS & G1 grism & 0.57-1.03 & 750 \\
\enddata
\tablecomments{The Gemini data were obtained through program GN-2008B-Q-21.
The data from SpeX \citep{ray03} were reduced with the
Spextool package \citep{cus04} and corrected for telluric absorption
\citep{vac03}.}
\end{deluxetable}

\begin{deluxetable}{llllllllll}
\tabletypesize{\scriptsize}
\tablewidth{0pt}
\tablecaption{Members of Taurus in Spectroscopic Sample\label{tab:new}}
\tablehead{
\colhead{} &
\colhead{} &
\colhead{Spectral} &
\colhead{$T_{\rm eff}$\tablenotemark{c}} &
\colhead{} &
\colhead{$L_{\rm bol}$} &
\colhead{Membership} &
\colhead{$W_{\lambda}$(H$\alpha$)} &
\colhead{Basis of} &
\colhead{} \\
\colhead{2MASS\tablenotemark{a}} &
\colhead{Other Names} &
\colhead{Type\tablenotemark{b}} &
\colhead{(K)} &
\colhead{$A_J$} &
\colhead{($L_\odot$)} &
\colhead{Evidence\tablenotemark{d}} &
\colhead{(\AA)} &
\colhead{Selection\tablenotemark{e}} &
\colhead{Night}}
\startdata
J04034997+2620382 & XEST 06-006 &   M5.25 &    3091 &       0 &   0.012 &     NaK,$\mu$ & 9$\pm$0.5 &     X-ray &       11 \\
J04144739+2803055 & XEST 20-066 &   M5.25 &    3091 &       0 &    0.12 &     NaK,$\mu$ & 8.5$\pm$0.5 &     X-ray &       21 \\
J04145234+2805598 & XEST 20-071 &   M3.25 &    3379 &    0.78 &    0.84 &  $A_V$,NaK,$\mu$ & 7.5$\pm$0.5 &     X-ray &       17 \\
J04153916+2818586 &    \nodata &   M3.75 &    3306 &    0.56 &    0.27 & ex,NaK,$A_V$ & 14$\pm$1 &  IR &        7 \\
J04154278+2909597 & IRAS 04125+2902 &   M1.25 &    3669 &    0.56 &    0.28 &   $A_V$,ex,$\mu$ & 2.3$\pm$0.3 &   IR &       45 \\
J04155799+2746175 &    \nodata &    M5.5 &    3058 &       0 &   0.049 & NaK,e,ex & 39$\pm$1 &    IR &       19 \\
J04181078+2519574 & V409 Tau,IRAS 04151+2512 &    M1.5 &    3632 &     1.3 &    0.53 & Li,ex,e,$\mu$ & 10$\pm$1 &    IR &       14 \\
J04193545+2827218 &      FR Tau,Haro6-4,XEST 23-076,XEST 24-063 &   M5.25 &    3091 &       0 &    0.10 & e,ex,NaK,$\mu$ & 67$\pm$5 &     IR &        6 \\
J04194657+2712552 &    [GKH94] 41 &   M7.5$\pm$1.5 &    2795 &     7.6 &    0.24 & $A_V$,ex,H$_2$O &      \nodata &   IR &       48 \\
J04201611+2821325 &    \nodata &    M6.5 &    2935 &       0 &  0.0072 & e,ex,NaK & 180$\pm$30 &     IR &        1 \\
J04202144+2813491 &    \nodata &   M1$\pm$2 &    3705 &    \nodata &      \nodata & $A_V$,e,ex &    $>$100 &     IR &       50 \\
J04202583+2819237 &  IRAS 04173+2812 &   mid-M &      \nodata &      \nodata &      \nodata &   e,ex &      \nodata &     IR &        9 \\
J04202606+2804089 &    \nodata &    M3.5 &    3342 &       0 &    0.15 &  ex,NaK & 5$\pm$0.3 &     IR &        4 \\
J04203918+2717317\tablenotemark{f} & XEST 16-045 &    M4.5 &    3198 &       0 &    0.16 &     NaK,$\mu$ & 8$\pm$0.5 &     X-ray &       21 \\
J04210934+2750368 &    \nodata &   M5.25 &    3091 &       0 &   0.079 &  ex,NaK & 20$\pm$2 &     IR &        6 \\
J04214013+2814224 & XEST 21-026 &   M5.75 &    3024 &       0 &   0.040 &     NaK,$\mu$ & 7$\pm$0.5 &     X-ray &       21 \\
J04221568+2657060\tablenotemark{f} & XEST 11-078,IRAS 04192+2650 &      M1 &    3705 &    0.28 &      \nodata & $A_V$,e,ex,$\mu$ & 14$\pm$1 &     X-ray &       21 \\
J04222404+2646258\tablenotemark{f} & XEST 11-087 &   M4.75 &    3161 &    0.31 &    0.12 &  $A_V$,NaK & 7.5$\pm$1 &     X-ray &       20 \\
J04233539+2503026 &     FU Tau A & M7.25 &    2838 &    0.56 &    0.19 & ex,H$_2$O,NaK,e,$\mu$ & 93$\pm$7 &    IR &    30,32 \\
J04233573+2502596 &     FU Tau B &   M9.25 &    2350 &       0 &  0.0025 & NaK,ex,e & $\sim$70 &    IR &       33 \\
J04293209+2430597\tablenotemark{g} &    \nodata &       ? &      \nodata &      \nodata &      \nodata &      ex &      \nodata &     IR &        9 \\
J04315968+1821305 &    LkHa 267 &   M1.5$\pm$0.5 &    3632 &     1.4 &    0.14 & $A_V$,e,ex &      \nodata &     IR &       49 \\
J04322415+2251083 &    \nodata &    M4.5 &    3198 &    0.49 &   0.090 & $A_V$,NaK,ex & 14$\pm$1 &    IR &       18 \\
J04324938+2253082 &    \nodata &   M4.25 &    3234 &    0.87 &    0.21 & $A_V$,ex,NaK & 18.5$\pm$1 &    IR &       22 \\
J04325119+1730092 &  LH 0429+17 & M8.25 &    2632 &       0 &  0.0033 & H$_2$O,NaK,$\mu$ & 17$\pm$10 &    classify &    30,33 \\
J04332621+2245293 & XEST 17-036 &      M4 &    3270 &     1.1 &    0.14 &  NaK,$A_V$ & 6$\pm$1 &     X-ray &       31 \\
J04333905+2227207 &    \nodata &   M1.75 &    3596 &    0.35 &      \nodata & $A_V$,e,ex & 23$\pm$1 &    IR &       17 \\
J04334171+1750402 &    \nodata & M4$\pm$0.5 &    3270 &    0.08 &   0.088 & $A_V$,ex,H$_2$O &      \nodata &     IR &       49 \\
J04334465+2615005 &    \nodata &   M4.75 &    3161 &    0.85 &    0.12 & e,ex,NaK,$A_V$ & 55$\pm$7 &   IR &        3 \\
J04335252+2256269\tablenotemark{f} & XEST 17-059 &   M5.75 &    3024 &       0 &    0.19 &     NaK,$\mu$ & 17$\pm$1 &     X-ray &       21 \\
J04345693+2258358\tablenotemark{f} & XEST 08-003 &    M1.5 &    3632 &    0.54 &    0.34 &      Li,$\mu$ & 2$\pm$0.2 &     X-ray &       13 \\
J04354203+2252226\tablenotemark{f} & XEST 08-033,XEST 09-023 &   M4.75 &    3161 &    0.49 &    0.12 &  $A_V$,NaK,$\mu$ & 14$\pm$1 &     X-ray &       20 \\
J04354733+2250216 &      HQ Tau,IRAS 04327+2244,XEST 09-026,XEST 08-037 &   K2$\pm$2 &    4900 &     1.2 &     4.6 &   Li,ex,$\mu$ & 2$\pm$0.4 &    IR &       10 \\
J04355209+2255039\tablenotemark{f} & XEST 08-047 &    M4.5 &    3198 &    0.56 &    0.13 &  $A_V$,NaK,$\mu$ & 6$\pm$0.5 &     X-ray &       21 \\
J04355286+2250585\tablenotemark{f} & XEST 08-049,XEST 09-033 &   M4.25 &    3234 &    0.35 &    0.14 &  NaK,e,$\mu$ & 25.5$\pm$1 &     X-ray &       27 \\
J04355892+2238353\tablenotemark{f} & XEST 09-042 &      M0 &    3850 &    0.11 &    0.71 &      Li,$\mu$ & 1.4$\pm$0.1 &     X-ray &       16 \\
J04373705+2331080 &    \nodata & L0$\pm$0.5 &    2200 &       0 & 0.0003 &  NaK &     $>$50 &     IR &       42 \\
J04414565+2301580 &    \nodata &   M4.5$\pm$0.5 &    3198 &    0.11 &    0.14 &     H$_2$O &      \nodata &   comp &       30 \\
J04455134+1555367 & IRAS 04429+1550 &   M2.5$\pm$0.5 &    3488 &    0.42 &    0.32 & Li,$A_V$,ex &      \nodata &     IR &       49 \\
J04554820+3030160 & XEST 26-052 &    M4.5 &    3198 &       0 &   0.045 &     NaK,$\mu$ & 6$\pm$0.5 &     X-ray &       17 \\
J04555605+3036209\tablenotemark{f} & JH433,XEST 26-062 &      M4 &    3270 &    0.53 &    0.28 & Li,e,ex & 17$\pm$2 &     X-ray &       24 \\
J04560118+3026348 & XEST 26-071 &   M3.5$\pm$0.5 &    3342 &    0.39 &    0.11 &   $A_V$,ex,$\mu$ &      \nodata &     IR &       49 \\
\enddata
\tablenotetext{a}{2MASS Point Source Catalog.}
\tablenotetext{b}{Uncertainties are $\pm0.25$ subclass unless noted otherwise.}
\tablenotetext{c}{Temperature scale from \citet{luh03ic}.}
\tablenotetext{d}{Membership in Taurus is indicated by $A_V\gtrsim1$ and
a position above the main sequence for the distance of Taurus (``$A_V$"),
strong emission lines (``e"), Na~I and K~I strengths intermediate
between those of dwarfs and giants (``NaK"),
strong Li absorption (``Li"), IR excess emission (``ex"),
the shape of the gravity-sensitive steam bands (``H$_2$O"),
or a proper motion that is similar to that of the known members of Taurus 
(``$\mu$").}
\tablenotetext{e}{Sources were selected for spectroscopy because
they were candidate companions to known Taurus members
\citep[``comp",][]{kra07} or they were identified as candidate members 
with {\it XMM-Newton} \citep[``X-ray",][]{sce07} or {\it Spitzer} 
(``IR", this work). Spectroscopy was also performed on a previously known
member to obtain a new spectral classification
\citep[``classify",][]{rei99}.}
\tablenotetext{f}{Independently confirmed as a member through spectroscopy by
\citet{sce08}.}
\tablenotetext{g}{Membership is uncertain because of the absence of 
spectroscopic evidence of youth.}
\end{deluxetable}

\begin{deluxetable}{llllll}
\tabletypesize{\scriptsize}
\tablewidth{0pt}
\tablecaption{Nonmembers in Spectroscopic Sample\label{tab:non}}
\tablehead{
\colhead{} &
\colhead{} &
\colhead{Spectral} &
\colhead{Basis of} &
\colhead{} \\
\colhead{ID\tablenotemark{a}} &
\colhead{Other Names} &
\colhead{Type\tablenotemark} &
\colhead{Selection\tablenotemark{b}} &
\colhead{Night}}
\startdata
2MASS J04042449+2611119 & XEST 06-041 &   M2.5V &     X-ray &       20 \\
2MASS J04124858+2749563 &    \nodata &   K0-K4 &    IR &       19 \\
2MASS J04141588+2818181 & XEST 20-045 &     $<$M4 &     X-ray &       30 \\
2MASS J04144294+2821105 & XEST 20-063 &     $<$M4 &     X-ray &       30 \\
2MASS J04164774+2408242 &    \nodata &      $<$M0 &    IR &       18 \\
2MASS J04170711+2408041 &    \nodata &  galaxy &    IR &       19 \\
2MASS J04180338+2440096 &    \nodata &      early A &     IR &       49 \\
2MASS J04180674+2904015 &    \nodata &      $<$M0 &    IR &       18 \\
2MASS J04182321+2519280 &    \nodata &  galaxy &    IR &       17 \\
2MASS J04190125+2837101 & XEST 23-062 &   giant &     X-ray &       22 \\
2MASS J04190689+2826090 & XEST 23-065,XEST 24-057 &     $<$M4 &     X-ray &       30 \\
2MASS J04191612+2750481 &    \nodata &  galaxy &     IR &        3 \\
2MASS J04214372+2647225 & XEST 11-035 &     $<$M0 &     X-ray &       16 \\
2MASS J04221295+2546598 &    \nodata &  galaxy &   IR &       30 \\
2MASS J04221918+2348005 &    \nodata &  galaxy &    IR &       12 \\
2MASS J04222559+2812332 & XEST 21-059 &     $<$M0 &     X-ray &       16 \\
2MASS J04222718+2659512 & XEST 11-088 &  M3.25V &     X-ray &       23 \\
2MASS J04223441+2457186 &    \nodata &  galaxy &    IR &       23 \\
2MASS J04224865+2823005 & XEST 21-073 &     M4V &     X-ray &       26 \\
2MASS J04263497+2608161 & XEST 02-005 &  galaxy &     X-ray &       39 \\
2MASS J04275871+2611062 & XEST 02-069 &     $<$M0 &     X-ray &       40 \\
2MASS J04285844+2436492 & XEST 13-002 &      M0 &     X-ray &       47 \\
2MASS J04292083+2742074 & IRAS 04262+2735 &   M5III &     IR &        4 \\
2MASS J04292887+2616483\tablenotemark{c} &    \nodata &   giant &   comp &       48 \\
2MASS J04293623+2634238 & LP358-731,XEST 15-034 &  M4.25V &     X-ray &       22 \\
2MASS J04301702+2622264 & LP358-352,XEST 15-075 &   M4.5V &     X-ray &       29 \\
2MASS J04302526+2602566 & XEST 14-034 &   M0.5V &     X-ray &       43 \\
2MASS J04302710+2807073 & IRAS 04273+2800 &  galaxy &     IR &        3 \\
2MASS J04304153+2430416 & XEST 13-036 &     $<$M0 &     X-ray &       16 \\
2MASS J04314419+2813170 &    \nodata &  galaxy &     IR &        8 \\
2MASS J04314634+2558404 & XEST 19-002 &  giant? &     X-ray &       17 \\
2MASS J04315860+1818408 & XEST 22-071 &     $<$M0 &     X-ray &       41 \\
2MASS J04322689+1818230 & XEST 22-111 &  M2.25V &     X-ray &       20 \\
2MASS J04322946+1814002 & XEST 22-114 &  galaxy &     X-ray &       20 \\
IRAC J04323521+2420213 &  XEST 03-026,XEST 04-001 &  galaxy &     IR &       44 \\
2MASS J04323605+2552225 & XEST 19-041 &   giant &     X-ray &       28 \\
2MASS J04323949+2427043 & XEST 03-028 &     $<$M4 &     X-ray &       30 \\
2MASS J04325921+2430403 & XEST 03-033,XEST 04-005 &  galaxy &     X-ray &       28 \\
2MASS J04332491+2559262 & XEST 19-083 &   giant &     X-ray &       23 \\
2MASS J04333301+2252521 & XEST 17-043 &   M4.5V &     X-ray &       36 \\
2MASS J04333746+2609550 &    \nodata &     $<$M4 &   comp &       30 \\
2MASS J04335562+2425016 & XEST 04-060 &     $<$M0 &     X-ray &       34 \\
2MASS J04341498+2826124 &    \nodata &   giant &     IR &        5 \\
2MASS J04343322+2602403 & JH87,XEST 18-059 &     $<$M0 &     X-ray &       38 \\
2MASS J04345164+2404426 & XEST 12-012 &   M3.5V &     X-ray &       35 \\
2MASS J04345973+2807017 &    \nodata &    M7V? &     IR &        9 \\
2MASS J04351316+2259205 & EZ Tau,LP358-739,XEST 08-014 &  M4.75V &     X-ray &       22 \\
2MASS J04353651+2304590 & XEST 08-027 &  M3.25V &     X-ray &       45 \\
2MASS J04354076+2411211 &    \nodata &     $<$M4 &   comp &       30 \\
2MASS J04400363+2553547 & [GKH94] 6,XEST 05-027 &   giant &     X-ray &       19 \\
2MASS J04403912+2540024 & XEST 07-005 &   M4.5V &     X-ray &       37 \\
IRAC J04412575+2543492 &  XEST 07-032 &  galaxy &     IR &       46 \\
2MASS J04415577+2302532 &    \nodata &  galaxy &   IR &       30 \\
2MASS J04420376+2519533 & XEST 10-016 &  giant? &     X-ray &       25 \\
2MASS J04455482+2408435 & IRAS 04428+2403 &  galaxy &     IR &        1 \\
2MASS J04455704+2440423 &    \nodata &   giant &     IR &        2 \\
2MASS J04553844+3031465 & JH427,XEST 26-031 &     $<$M0 &     X-ray &       15 \\
2MASS J04562935+3036115 & XEST 26-135 &      $<$M0 &     X-ray &       17 \\
\enddata
\tablenotetext{a}{Sources that are not in the 2MASS Point Source Catalog
have been assigned coordinate-based identifications using IRAC images.}
\tablenotetext{b}{Sources were selected for spectroscopy because
they were candidate companions to known Taurus members
\citep[``comp",][]{kra07} or they were identified as candidate members 
with {\it XMM-Newton} \citep[``X-ray",][]{sce07} or {\it Spitzer} 
(``IR", this work).}
\tablenotetext{c}{Mistakenly identified as FW Tau B by \citet{har05}.
\citet{wg01} classified it as a probable background star based on the presence
of H$\alpha$ absorption.}
\end{deluxetable}

\begin{deluxetable}{ll}
\tablewidth{0pt}
\tablecaption{Measurements of Li Absorption\label{tab:li}}
\tablehead{
\colhead{} &
\colhead{$W_{\lambda}$} \\
\colhead{Name} &
\colhead{(\AA)}}
\startdata
HQ Tau & 0.4$\pm$0.05 \\
XEST 09-042 & 0.55$\pm$0.05 \\
XEST 08-003 & 0.6$\pm$0.05 \\
V409 Tau & 0.6$\pm$0.1 \\
XEST 26-062 & 0.47$\pm$0.05 \\
\enddata
\end{deluxetable}

\begin{deluxetable}{llllll}
\tabletypesize{\scriptsize}
\tablewidth{0pt}
\tablecaption{IRAC Photometry for Nonmembers\label{tab:nonirac}}
\tablehead{
\colhead{ID} &
\colhead{[3.6]} &
\colhead{[4.5]} &
\colhead{[5.8]} &
\colhead{[8.0]} &
\colhead{Date}
}
\startdata
2MASS J04124858+2749563 & 11.40$\pm$0.02 & 10.87$\pm$0.02 & 10.32$\pm$0.03 &  9.14$\pm$0.03 & 2007 Mar 29 \\
2MASS J04141588+2818181 & 12.36$\pm$0.02 & 12.11$\pm$0.02 & 12.13$\pm$0.04 & 12.10$\pm$0.04 & 2005 Feb 19 \\
                  & 12.21$\pm$0.02 & 12.07$\pm$0.02 & 12.10$\pm$0.04 & 12.00$\pm$0.04 & 2007 Mar 29 \\
2MASS J04144294+2821105 & 11.75$\pm$0.02 & 11.62$\pm$0.02 & 11.54$\pm$0.03 & 11.56$\pm$0.04 & 2005 Feb 19 \\
                  & 11.70$\pm$0.02 & 11.77$\pm$0.02 & 11.49$\pm$0.03 & 11.46$\pm$0.04 & 2007 Mar 29 \\
2MASS J04164774+2408242 & 13.84$\pm$0.03 & 13.36$\pm$0.03 & 13.12$\pm$0.05 & 12.26$\pm$0.04 & 2007 Mar 30 \\
2MASS J04170711+2408041 & 13.93$\pm$0.03 & 13.72$\pm$0.03 & 13.53$\pm$0.06 & 12.63$\pm$0.06 & 2007 Mar 30 \\
2MASS J04180338+2440096 & 10.04$\pm$0.02 & 10.03$\pm$0.02 & 10.03$\pm$0.03 &  9.83$\pm$0.03 & 2007 Mar 30 \\
2MASS J04180674+2904015 & 13.74$\pm$0.03 & 13.48$\pm$0.03 & 13.51$\pm$0.06 & 12.61$\pm$0.06 & 2007 Mar 29 \\
2MASS J04182321+2519280 & 11.76$\pm$0.02 & 11.06$\pm$0.02 & 10.45$\pm$0.03 &  9.62$\pm$0.03 & 2007 Mar 30 \\
2MASS J04190125+2837101 &        out & 10.69$\pm$0.02 &        out & 10.65$\pm$0.03 & 2005 Feb 19 \\
                  & 10.68$\pm$0.02 & 10.67$\pm$0.02 & 10.68$\pm$0.03 & 10.65$\pm$0.03 & 2007 Mar 29 \\
2MASS J04190689+2826090 & 12.52$\pm$0.02 & 12.42$\pm$0.02 & 12.38$\pm$0.04 & 12.31$\pm$0.04 & 2005 Feb 19 \\
                  & 12.55$\pm$0.02 & 12.43$\pm$0.02 & 12.39$\pm$0.04 & 12.25$\pm$0.04 & 2005 Feb 21 \\
2MASS J04191612+2750481 & 11.84$\pm$0.02 & 11.15$\pm$0.02 & 10.48$\pm$0.03 &  9.39$\pm$0.03 & 2005 Feb 21 \\
2MASS J04214372+2647225 &  8.81$\pm$0.02 &  8.85$\pm$0.02 &  8.77$\pm$0.03 &  8.77$\pm$0.03 & 2007 Oct 17 \\
2MASS J04221295+2546598 & 12.60$\pm$0.02 & 11.81$\pm$0.02 & 11.16$\pm$0.03 & 10.37$\pm$0.03 & 2006 Sep 28 \\
                  & 12.54$\pm$0.02 & 11.80$\pm$0.02 & 11.17$\pm$0.03 & 10.32$\pm$0.03 & 2007 Mar 30 \\
2MASS J04221918+2348005 & 14.08$\pm$0.03 & 13.59$\pm$0.03 & 13.42$\pm$0.06 & 12.25$\pm$0.05 & 2007 Mar 30 \\
2MASS J04222559+2812332 &  8.54$\pm$0.02 &  8.54$\pm$0.02 &  8.55$\pm$0.03 &  8.50$\pm$0.03 & 2005 Feb 21 \\
2MASS J04222718+2659512 & 13.92$\pm$0.02 & 13.75$\pm$0.03 & 13.77$\pm$0.07 & \nodata & 2007 Oct 17 \\
2MASS J04223441+2457186 & 14.77$\pm$0.03 & 14.13$\pm$0.03 & 13.64$\pm$0.06 & 12.87$\pm$0.06 & 2007 Mar 30 \\
2MASS J04224865+2823005 & 13.17$\pm$0.02 & 13.07$\pm$0.02 & 13.03$\pm$0.05 & 13.23$\pm$0.07 & 2005 Feb 21 \\
2MASS J04263497+2608161 & 13.09$\pm$0.02 & 12.55$\pm$0.02 & 11.66$\pm$0.03 &  9.63$\pm$0.03 & 2005 Feb 22 \\
2MASS J04275871+2611062 &  9.77$\pm$0.02 &        out &  9.71$\pm$0.03 &     out & 2004 Mar 07 \\
                  &  9.80$\pm$0.02 &  9.85$\pm$0.02 &  9.84$\pm$0.03 &  9.75$\pm$0.03 & 2005 Feb 22 \\
2MASS J04285844+2436492 & 10.63$\pm$0.02 & 10.54$\pm$0.02 & 10.47$\pm$0.03 & 10.47$\pm$0.03 & 2005 Feb 20 \\
2MASS J04292083+2742074 &  6.66$\pm$0.02 &  6.52$\pm$0.02 &  6.18$\pm$0.03 &  4.98$\pm$0.03 & 2005 Feb 24 \\
2MASS J04292887+2616483 &  9.22$\pm$0.02 &  9.21$\pm$0.02 &  9.12$\pm$0.03 &  9.08$\pm$0.03 & 2004 Mar 07 \\
                  &  9.21$\pm$0.02 &  9.24$\pm$0.02 &  9.10$\pm$0.03 &  9.12$\pm$0.03 & 2005 Feb 24 \\
                  &  9.22$\pm$0.02 &  9.15$\pm$0.02 &  9.15$\pm$0.03 &  9.07$\pm$0.03 & 2006 Mar 25 \\
2MASS J04293623+2634238 &        out & 10.30$\pm$0.02 &        out & 10.23$\pm$0.03 & 2004 Mar 07 \\
                  & 10.38$\pm$0.02 & 10.33$\pm$0.02 & 10.26$\pm$0.03 & 10.27$\pm$0.03 & 2005 Feb 24 \\
2MASS J04301702+2622264 & 10.63$\pm$0.02 & 10.57$\pm$0.02 & 10.58$\pm$0.03 & 10.52$\pm$0.03 & 2005 Feb 24 \\
2MASS J04302526+2602566 & 10.04$\pm$0.02 & 10.01$\pm$0.02 & 10.04$\pm$0.03 &  9.97$\pm$0.03 & 2005 Feb 24 \\
2MASS J04302710+2807073 & 10.82$\pm$0.02 &  9.68$\pm$0.02 &  8.70$\pm$0.03 &  7.61$\pm$0.03 & 2005 Feb 24 \\
2MASS J04304153+2430416 &  9.20$\pm$0.02 &  9.21$\pm$0.02 &  9.17$\pm$0.03 &  9.19$\pm$0.03 & 2005 Feb 24 \\
2MASS J04314419+2813170 & 14.04$\pm$0.03 & 13.92$\pm$0.03 & 13.97$\pm$0.07 & 12.83$\pm$0.06 & 2005 Feb 20 \\
                  & 13.96$\pm$0.03 & 13.87$\pm$0.03 & 13.71$\pm$0.08 & 12.88$\pm$0.06 & 2005 Feb 24 \\
2MASS J04314634+2558404 & 12.12$\pm$0.02 & 12.12$\pm$0.02 & 12.07$\pm$0.04 & 12.06$\pm$0.03 & 2005 Feb 24 \\
2MASS J04315860+1818408 &  9.10$\pm$0.02 &  9.09$\pm$0.02 &  9.14$\pm$0.03 &  9.03$\pm$0.03 & 2004 Oct 07 \\
2MASS J04322689+1818230 & 11.46$\pm$0.02 & 11.45$\pm$0.02 & 11.51$\pm$0.03 & 11.36$\pm$0.03 & 2004 Oct 07 \\
                  & 11.47$\pm$0.02 &        out & 11.38$\pm$0.03 &     out & 2005 Feb 19 \\
2MASS J04322946+1814002 & 12.68$\pm$0.02 & 12.54$\pm$0.02 & 12.17$\pm$0.04 & 10.90$\pm$0.03 & 2004 Oct 07 \\
IRAC J04323521+2420213 & 13.58$\pm$0.03 & 12.56$\pm$0.02 & 11.73$\pm$0.03 & 10.76$\pm$0.03 & 2004 Oct 07 \\
                  & 13.54$\pm$0.02 & 12.54$\pm$0.02 & 11.69$\pm$0.04 & 10.77$\pm$0.03 & 2005 Feb 20 \\
                  & 13.55$\pm$0.03 & 12.65$\pm$0.03 & 11.72$\pm$0.04 & 10.80$\pm$0.03 & 2005 Feb 24 \\
2MASS J04323605+2552225 & 13.84$\pm$0.02 & 13.80$\pm$0.03 & 13.74$\pm$0.07 & 13.52$\pm$0.09 & 2004 Mar 07 \\
                  & 13.89$\pm$0.02 & 13.80$\pm$0.02 & 13.84$\pm$0.06 & 13.48$\pm$0.07 & 2005 Feb 24 \\
2MASS J04323949+2427043 & 11.94$\pm$0.02 & 11.86$\pm$0.02 & 11.76$\pm$0.03 & 11.77$\pm$0.03 & 2004 Oct 07 \\
                  & 11.97$\pm$0.02 & 11.82$\pm$0.02 & 11.68$\pm$0.03 & 11.71$\pm$0.04 & 2005 Feb 24 \\
                  & 11.93$\pm$0.02 & 11.80$\pm$0.02 & 11.77$\pm$0.03 & 11.69$\pm$0.03 & 2006 Mar 26 \\
2MASS J04325921+2430403 & 14.00$\pm$0.02 &        out & 13.54$\pm$0.05 &     out & 2004 Oct 07 \\
                  & 13.98$\pm$0.03 & 13.74$\pm$0.03 & 13.44$\pm$0.06 & 11.36$\pm$0.04 & 2005 Feb 24 \\
2MASS J04332491+2559262 & 13.08$\pm$0.02 & 13.01$\pm$0.02 & 13.10$\pm$0.05 & 12.98$\pm$0.05 & 2005 Feb 24 \\
2MASS J04333301+2252521 & 14.07$\pm$0.03 &        out & 13.81$\pm$0.07 &     out & 2005 Feb 24 \\
                  & 13.98$\pm$0.02 & 14.02$\pm$0.03 & 14.17$\pm$0.10 & \nodata & 2007 Apr 03 \\
2MASS J04333746+2609550 & 12.82$\pm$0.04 & 12.63$\pm$0.08 & 12.58$\pm$0.05 & \nodata & 2004 Mar 07 \\
                  & 12.78$\pm$0.03 & 12.71$\pm$0.05 & 12.61$\pm$0.05 & \nodata & 2005 Feb 24 \\
2MASS J04335562+2425016 &  8.78$\pm$0.02 &  8.83$\pm$0.02 &  8.85$\pm$0.03 &  8.76$\pm$0.03 & 2005 Feb 24 \\
2MASS J04341498+2826124 & 13.83$\pm$0.02 & 13.75$\pm$0.03 & 13.73$\pm$0.06 & 13.11$\pm$0.06 & 2005 Feb 20 \\
2MASS J04343322+2602403 & 10.06$\pm$0.02 & 10.03$\pm$0.02 &  9.97$\pm$0.03 & 10.01$\pm$0.03 & 2005 Feb 22 \\
2MASS J04345164+2404426 & 11.63$\pm$0.02 & 11.54$\pm$0.02 & 11.50$\pm$0.03 & 11.49$\pm$0.03 & 2006 Mar 25 \\
                  & 11.64$\pm$0.02 & 11.55$\pm$0.02 & 11.58$\pm$0.03 & 11.52$\pm$0.03 & 2007 Oct 16 \\
2MASS J04345973+2807017 & 14.34$\pm$0.03 & 13.95$\pm$0.03 & 13.71$\pm$0.06 & 13.15$\pm$0.1 & 2005 Feb 20 \\
                  & 14.21$\pm$0.03 & 13.91$\pm$0.03 & 13.79$\pm$0.07 & 13.04$\pm$0.1 & 2007 Oct 16 \\
2MASS J04351316+2259205 &        out &  9.91$\pm$0.02 &        out &  9.84$\pm$0.03 & 2005 Feb 20 \\
                  &  9.91$\pm$0.02 &        out &  9.84$\pm$0.03 &     out & 2005 Feb 24 \\
                  &  9.95$\pm$0.02 &  9.90$\pm$0.02 &  9.86$\pm$0.03 &  9.84$\pm$0.03 & 2007 Apr 03 \\
2MASS J04353651+2304590 & 14.52$\pm$0.03 & 14.39$\pm$0.03 & 14.55$\pm$0.12 & \nodata & 2005 Feb 21 \\
                  &        out & 14.40$\pm$0.04 &        out & \nodata & 2007 Apr 03 \\
2MASS J04354076+2411211 & 12.59$\pm$0.03 & 12.34$\pm$0.04 &    \nodata & \nodata & 2005 Feb 20 \\
                  & 12.60$\pm$0.03 & 12.29$\pm$0.05 &    \nodata & \nodata & 2005 Feb 21 \\
                  & 12.52$\pm$0.02 & 12.26$\pm$0.04 &    \nodata & \nodata & 2006 Mar 25 \\
2MASS J04400363+2553547 &  7.55$\pm$0.02 &        out &  7.33$\pm$0.03 &     out & 2004 Mar 07 \\
                  &  7.54$\pm$0.02 &  7.42$\pm$0.02 &  7.39$\pm$0.03 &  7.34$\pm$0.03 & 2005 Feb 22 \\
2MASS J04403912+2540024 & 13.64$\pm$0.02 & 13.51$\pm$0.03 & 13.49$\pm$0.06 & 13.54$\pm$0.09 & 2007 Oct 16 \\
IRAC J04412575+2543492 & 13.73$\pm$0.02 & 12.39$\pm$0.02 & 11.22$\pm$0.03 & 10.22$\pm$0.03 & 2004 Oct 08 \\
                  & 13.76$\pm$0.02 & 12.38$\pm$0.02 & 11.21$\pm$0.03 & 10.25$\pm$0.03 & 2005 Feb 23 \\
2MASS J04415577+2302532 & 12.86$\pm$0.02 & 12.24$\pm$0.02 & 11.64$\pm$0.03 & 10.40$\pm$0.03 & 2007 Mar 28 \\
2MASS J04420376+2519533 & 13.22$\pm$0.02 & 13.16$\pm$0.02 & 13.17$\pm$0.05 & 12.99$\pm$0.06 & 2005 Feb 23 \\
2MASS J04455482+2408435 & 12.16$\pm$0.02 & 11.46$\pm$0.02 & 10.44$\pm$0.03 &  8.67$\pm$0.03 & 2007 Oct 16 \\
2MASS J04455704+2440423 & 13.03$\pm$0.02 & 12.91$\pm$0.02 & 12.63$\pm$0.04 & 12.08$\pm$0.04 & 2005 Feb 24 \\
2MASS J04553844+3031465 & 10.03$\pm$0.02 & 10.09$\pm$0.02 & 10.00$\pm$0.03 & 10.04$\pm$0.03 & 2004 Feb 14 \\
                  & 10.06$\pm$0.02 &        out & 10.02$\pm$0.03 &     out & 2005 Feb 20 \\
\enddata
\tablecomments{Entries of ``$\cdots$" and ``out"
indicate measurements that are absent because of non-detection
and a position outside the field of view of the camera, respectively.}
\end{deluxetable}

\begin{deluxetable}{lll}
\tabletypesize{\scriptsize}
\tablewidth{0pt}
\tablecaption{MIPS 24~\micron\ Photometry for Nonmembers\label{tab:nonmips}}
\tablehead{
\colhead{ID} &
\colhead{[24]} &
\colhead{Date}
}
\startdata
2MASS J04124858+2749563 &  6.46$\pm$0.04 & 2007 Feb 27 \\
2MASS J04164774+2408242 &  8.98$\pm$0.11 & 2007 Feb 26 \\
2MASS J04180338+2440096 &  7.20$\pm$0.05 & 2007 Feb 23 \\
2MASS J04182321+2519280 &  6.17$\pm$0.04 & 2007 Feb 23 \\
2MASS J04191612+2750481 &  5.25$\pm$0.04 & 2005 Feb 28 \\
                  &  5.28$\pm$0.04 & 2007 Oct 28 \\
2MASS J04214372+2647225 &  8.47$\pm$0.11 & 2007 Feb 28 \\
                  &  8.99$\pm$0.10 & 2007 Oct 28 \\
2MASS J04221295+2546598 &  7.25$\pm$0.04 & 2004 Sep 25 \\
                  &  7.20$\pm$0.05 & 2007 Feb 28 \\
2MASS J04221918+2348005 &  8.67$\pm$0.18 & 2007 Feb 28 \\
2MASS J04222559+2812332 &  8.39$\pm$0.09 & 2005 Feb 27 \\
2MASS J04263497+2608161 &  5.34$\pm$0.04 & 2005 Feb 28 \\
                  &  5.38$\pm$0.04 & 2005 Mar 01 \\
2MASS J04292083+2742074 &  3.05$\pm$0.04 & 2005 Mar 02 \\
2MASS J04302710+2807073 &  3.68$\pm$0.04 & 2005 Mar 01 \\
2MASS J04304153+2430416 &  8.90$\pm$0.10 & 2005 Feb 26 \\
                  &  8.85$\pm$0.12 & 2005 Mar 02 \\
2MASS J04315860+1818408 &  8.89$\pm$0.14 & 2004 Feb 20 \\
                  &  9.10$\pm$0.08 & 2004 Sep 25 \\
                  &  9.15$\pm$0.13 & 2006 Feb 19 \\
2MASS J04322946+1814002 &  7.38$\pm$0.05 & 2004 Feb 20 \\
                  &  7.40$\pm$0.04 & 2004 Sep 25 \\
                  &  7.42$\pm$0.04 & 2006 Feb 19 \\
IRAC J04323521+2420213 &  7.06$\pm$0.05 & 2004 Sep 25 \\
                  &  7.07$\pm$0.05 & 2005 Mar 01 \\
                  &  7.10$\pm$0.05 & 2007 Sep 23 \\
2MASS J04325921+2430403 &  8.30$\pm$0.09 & 2004 Sep 25 \\
                  &  8.49$\pm$0.16 & 2005 Feb 26 \\
                  &  8.81$\pm$0.22 & 2005 Mar 02 \\
                  &  8.59$\pm$0.07 & 2007 Sep 24 \\
2MASS J04335562+2425016 &  8.40$\pm$0.08 & 2005 Mar 01 \\
2MASS J04400363+2553547 &  7.33$\pm$0.05 & 2005 Mar 04 \\
IRAC J04412575+2543492 &  6.51$\pm$0.04 & 2004 Sep 25 \\
                  &  6.63$\pm$0.04 & 2005 Feb 26 \\
                  &  6.71$\pm$0.05 & 2005 Feb 28 \\
                  &  6.52$\pm$0.04 & 2007 Sep 25 \\
2MASS J04455482+2408435 &  5.00$\pm$0.04 & 2005 Feb 28 \\
\enddata
\end{deluxetable}

\begin{deluxetable}{lllllll}
\tabletypesize{\scriptsize}
\tablewidth{0pt}
\tablecaption{IRAC Photometry for Young Cool Field Dwarfs\label{tab:ldwarfs}}
\tablehead{
\colhead{} &
\colhead{Spectral} &
\colhead{} &
\colhead{} &
\colhead{} &
\colhead{} &
\colhead{} \\
\colhead{ID} &
\colhead{Type\tablenotemark{a}} &
\colhead{[3.6]} &
\colhead{[4.5]} &
\colhead{[5.8]} &
\colhead{[8.0]} &
\colhead{Date}
}
\startdata
2MASS J00332386$-$1521309 &  L4 & 12.54$\pm$0.02 & 12.48$\pm$0.02 & 12.21$\pm$0.03 & 12.04$\pm$0.03 & 2007 Aug 10 \\
2MASS J01415823$-$4633574 &  L0 & 12.36$\pm$0.02 & 12.17$\pm$0.02 & 11.95$\pm$0.03 & 11.68$\pm$0.03 & 2006 Aug 12 \\
2MASS J02411151$-$0326587 &  L0 & 13.39$\pm$0.02 & 13.24$\pm$0.02 & 13.04$\pm$0.03 & 12.77$\pm$0.03 & 2007 Sep 12 \\
2MASSI J0253597+320637 &  M7 & 12.17$\pm$0.02 & 12.12$\pm$0.02 & 12.12$\pm$0.03 & 12.01$\pm$0.03 & 2007 Sep 7 \\
2MASS J03231002$-$4631237 &  L0 & 12.84$\pm$0.02 & 12.68$\pm$0.02 & 12.48$\pm$0.03 & 12.16$\pm$0.03 & 2007 Aug 10 \\
2MASS J03572695$-$4417305 &  L0 & 12.21$\pm$0.02 & 12.08$\pm$0.02 & 11.86$\pm$0.03 & 11.65$\pm$0.03 & 2007 Aug 10 \\
SDSS J044337.61+000205.1 &  M9 & 10.55$\pm$0.02 & 10.45$\pm$0.02 & 10.35$\pm$0.03 & 10.22$\pm$0.03 & 2007 Oct 16 \\
2MASS J05012406$-$0010452 &  L4 & 11.77$\pm$0.02 & 11.52$\pm$0.02 & 11.22$\pm$0.03 & 11.03$\pm$0.03 & 2008 Mar 9 \\
2MASSI J0608528$-$275358 & M8.5 & 11.75$\pm$0.02 & 11.62$\pm$0.02 & 11.52$\pm$0.03 & 11.44$\pm$0.03 & 2007 Oct 16 \\
2MASSI J1615425+495321 &  L4 & 12.91$\pm$0.02 & 12.61$\pm$0.02 & 12.28$\pm$0.03 & 12.15$\pm$0.03 & 2007 Jul 2 \\
2MASSI J1726000+153819 &  L3 & 12.76$\pm$0.02 & 12.64$\pm$0.02 & 12.41$\pm$0.03 & 12.20$\pm$0.03 & 2007 Sep 7 \\
2MASSW J2208136+292121 &  L3 & 13.08$\pm$0.02 & 12.89$\pm$0.02 & 12.62$\pm$0.03 & 12.33$\pm$0.03 & 2007 Jul 3 \\
2MASS J22134491$-$2136079 &  L0 & 12.99$\pm$0.02 & 12.83$\pm$0.02 & 12.58$\pm$0.03 & 12.37$\pm$0.03 & 2007 Jun 29 \\
\enddata
\tablenotetext{a}{\citet{kir06,kir08} and \citet{cru07,cru09}.}
\end{deluxetable}

\begin{deluxetable}{lllrrlrlllll}
\tabletypesize{\scriptsize}
\tablewidth{0pt}
\tablecaption{Median Positions, Proper Motions, and Velocities of Taurus Groups\label{tab:pm}}
\tablehead{
\colhead{Group\tablenotemark{a}} &
\colhead{$\alpha$(J2000)} &
\colhead{$\delta$(J2000)} &
\colhead{$\mu_{\alpha}$\tablenotemark{b}} &
\colhead{$\mu_{\delta}$\tablenotemark{b}} &
\colhead{N$_{\mu}$\tablenotemark{c}} &
\colhead{${\rm RV}_{star}$\tablenotemark{d}} &
\colhead{${\rm RV}_{gas}$\tablenotemark{d}} &
\colhead{$U$\tablenotemark{e}} &
\colhead{$V$\tablenotemark{e}} &
\colhead{$W$\tablenotemark{e}} &
\colhead{Cloud}}
\startdata
I   & 4 14 26 & 28 12 00 &  $+6.9$ & $-22.3$ & 63  & 12.3\,$\pm$\,2.5 & 15.5  & $-$15 & $-$11 & $-$11 & B209\\
II  & 4 19 24 & 28 20 20 &  $+6.0$ & $-26.8$ & 101 & 14.9\,$\pm$\,0.6 & 15.6  & $-$15 & $-$13 & $-$13 & L1495E\\
III & 4 40 55 & 25 40 50 &  $+4.5$ & $-21.3$ & 57  & 16.3\,$\pm$\,1.0 & 15.9  & $-$15 & $-$11 & $-$10 & L1527\\
IV  & 4 32 36 & 24 21 40 &  $+5.5$ & $-21.9$ & 79  & 16.8\,$\pm$\,0.7 & 16.6 & $-$15 & $-$12 & $-$11 & L1529\\
V   & 4 35 55 & 22 52 50 &  $+6.7$ & $-17.7$ & 88  & 15.7\,$\pm$\,1.5 & 16.4 & $-$15 & $-$11 & $-$9 & L1536\\
VI  & 4 32 22 & 18 10 50 & $+10.0$ & $-17.6$ & 87  & 17.9\,$\pm$\,0.3 & 18.5 & $-$17 & $-$13 & $-$8 & L1551\\
VII & 4 21 58 & 19 32 20 & $+12.2$ & $-12.7$ & 12  & 17.6\,$\pm$\,1.0 & \nodata     & $-$17 & $-$11 & $-$6 & NGC1554\\ 
VIII & 4 28 41 & 26 19 10 &  $+8.6$ & $-22.0$ & 111 & 14.9\,$\pm$\,0.6 & 16.4 & $-$16 & $-$13 & $-$10 & B217\\
IX  & 4 47 00 & 17 00 40 &  $+3.2$ & $-16.1$ & 12  & 22.5\,$\pm$\,1.1 & 20.6 & $-$18 & $-$11 & $-$11 & L1558\\
X   & 4 55 29 & 30 30 40 &  $+3.9$ & $-23.4$ & 52  & 14.8\,$\pm$\,0.5 & 14.9 & $-$15 & $-$12 & $-$10 & L1517\\
XI  & 5 06 00 & 24 58 10 &  $+0.9$ & $-17.6$ & 24  & 16.8\,$\pm$\,3.8 & 17.9 & $-$17 & $-$9 & $-$9 & L1544\\
all of Taurus & 4 32 10 & 25 51 40 &  $+6.1$ & $-21.0$ & 757  & 16.3\,$\pm$\,0.3 & 16.4 & $-$16 & $-$11 & $-$10 & \nodata
\enddata
\tablecomments{Proper motions ($\mu_{\alpha}$, $\mu_{\delta}$) and velocities
(RV, $U$, $V$, $W$) have units of mas~yr$^{-1}$ and km\,s$^{-1}$, respectively.}
\tablenotetext{a}{Group names I through VI are from \citet{gom93}.
We have defined the designations for the remaining groups.}
\tablenotetext{b}{Uncertainties in the median $\mu_{\alpha}$ and $\mu_{\delta}$ values
for each group are $\sim$1 mas~yr$^{-1}$ for groups I, II, IV, VI,
VIII, X, and XI, and $\sim$2 mas~yr$^{-1}$ for groups III, V, VII, and
IX.}
\tablenotetext{c}{Number of proper motion measurements included in the
calculation of the median motion.}
\tablenotetext{d}{Heliocentric velocities.}
\tablenotetext{e}{Galactic Cartesian velocities.}
\end{deluxetable}

\clearpage

\begin{thebibliography}{}


\bibitem[Allen \& Strom(1995)]{as95}
Allen, L. E., \& Strom, K. M. 1995, \aj, 109, 1379

\bibitem[Andr\'e et al.(1999)]{and99}
Andr\'e, P., Motte, F., \& Bacmann, A. 1999, \apj, 513, L57

\bibitem[Appenzeller et al.(1988)]{app88} Appenzeller, I., 
Reitermann, A., \& Stahl, O. 1988, \pasp, 100, 815 

\bibitem[Baraffe et al.(1998)]{bar98}
Baraffe, I., Chabrier, G., Allard, F., \& Hauschildt, P. H. 1998, \aap, 337, 403

\bibitem[Barbier-Brossat et al.(1994)]{bbf94} 
Barbier-Brossat, M., Petit, M., \& Figon, P. 1994, \aaps, 108, 603 

\bibitem[Barbier-Brossat \& Figon(2000)]{bbf00} 
Barbier-Brossat, M., \& Figon, P. 2000, \aaps, 142, 217 

\bibitem[Bertout et al.(2007)]{ber07} 
Bertout, C., Siess, L., \& Cabrit, S. 2007, \aap, 473, L21

\bibitem[Bourke et al.(2006)]{bou06}
Bourke, T. L., et al. 2006, \apj, 649, L37

\bibitem[Brice\~no et al.(2002)]{bri02}
Brice\~{n}o, C., Luhman, K. L., Hartmann, L., Stauffer, J. R., \& Kirkpatrick,
J. D. 2002, \apj, 580, 317

\bibitem[Brown et al.(2008)]{bro08}
Brown, J. M., Blake, G. A., Qi, C., Dullemond, C. P., \& Wilner, D. J. 2008,
\apj, 675, L109

\bibitem[Burrows et al.(1996)]{bur96}
Burrows, C. J., et al. 1996, \apj, 473, 437

\bibitem[Calvet et al.(2002)]{cal02}
Calvet, N., D'Alessio, P., Hartmann, L., Wilner, D., Walsh, A., \& Sitko, M.
2002, \apj, 568, 1008

\bibitem[Calvet et al.(2005)]{cal05} 
Calvet, N., et al. 2005, \apj, 630, L185

\bibitem[Chabrier et al.(2000)]{cha00} 
Chabrier, G., Baraffe, I., Allard, F., \& Hauschildt, P. 2000, \apj, 542, L119

\bibitem[Cruz et al.(2009)]{cru09} 
Cruz, K. L., Kirkpatrick, J. D., \& Burgasser, A. J. 2009, \aj, 137, 3345

\bibitem[Cruz et al.(2007)]{cru07}
Cruz, K. L., et al. 2007, \aj, 133, 439

\bibitem[Cushing et al.(2005)]{cus05} 
Cushing, M. C., Rayner, J. T., \& Vacca, W. D. 2005, \apj, 623, 1115

\bibitem[Cushing et al.(2004)]{cus04} 
Cushing, M. C., Vacca, W. D., \& Rayner, J. T. 2004, \pasp, 116, 362

\bibitem[D'Alessio et al.(2005)]{dal05} 
D'Alessio, P., et al. 2005, \apj, 621, 461

\bibitem[Dobashi et al.(2005)]{dob05} 
Dobashi, K., Uehara, H., Kandori, R., Sakurai, T., Kaiden, M., Umemoto, T., 
\& Sato, F. 2005, \pasj, 57, 1

\bibitem[Ducourant et al.(2005)]{duc05} 
Ducourant, C., Teixeira, R., P{\'e}ri{\'e}, J. P., Lecampion, J. F.,
Guibert, J., \& Sartori, M. J. 2005, \aap, 438, 769

\bibitem[Dutrey et al.(2008)]{dut08} 
Dutrey, A., et al. 2008, \aap, 490, L15

\bibitem[Espaillat et al.(2007a)]{esp07a} 
Espaillat, C., et al. 2007a, \apj, 664, L111

\bibitem[Espaillat et al.(2007b)]{esp07b} 
Espaillat, C., et al. 2007b, \apj, 670, L135

\bibitem[Faherty et al.(2009)]{fah09} 
Faherty, J. K., Burgasser, A. J., Cruz, K. L., Shara, M. M., Walter, F. M., 
\& Gelino, C. R. 2009, \aj, 137, 1

\bibitem[Fazio et al.(2004)]{faz04} 
Fazio, G. G., et al. 2004, \apjs, 154, 10

\bibitem[Finkenzeller \& Jankovics(1984)]{fin84}
Finkenzeller, U., \& Jankovics, I. 1984, \aaps, 57, 285

\bibitem[Flaherty et al.(2007)]{fla07} 
Flaherty, K. M., Pipher, J. L., Megeath, S. T., Winston, E. M., 
Gutermuth, R. A., Muzerolle, J., Allen, L. E., \& Fazio, G. G. 
2007, \apj, 663, 1069

\bibitem[Furlan et al.(2006)]{fur06}
Furlan, E., et al. 2006, \apjs, 165, 568

\bibitem[Furlan et al.(2007)]{fur07} 
Furlan, E., et al. 2007, \apj, 644, 1176

\bibitem[Furlan et al.(2008)]{fur08} 
Furlan, E., et al. 2008, \apjs, 176, 184

\bibitem[Gahm et al.(1999)]{gah99} 
Gahm, G. F., Petrov, P. P., Duemmler, R., Gameiro, J. F., \&
Lago, M. T. V. T. 1999, \aap, 352, L95

\bibitem[Gomez et al.(1993)]{gom93} 
Gomez, M., Hartmann, L., Kenyon, S. J., \& Hewett, R. 1993, \aj, 105, 1927

\bibitem[Goodwin et al.(2004)]{goo04} 
Goodwin, S. P., Whitworth, A. P., \& Ward-Thompson, D. 2004, \aap, 419, 543

\bibitem[Grosso et al.(2007)]{gro07} 
Grosso, N., et al. 2007, \aap, 468, 391

\bibitem[G\"udel et al.(2007)]{gud07} 
G\"udel, M., et al. 2007, \aap, 468, 353

\bibitem[Guieu et al.(2006)]{gui06} 
Guieu, S., Dougados, C., Monin, J.-L., Magnier, E. \& Mart{\'\i}n, E. L. 2006,
\aap, 446, 485

\bibitem[Guieu et al.(2007)]{gui07}
Guieu, S., et al. 2007, \aap, 465, 855

\bibitem[Hambly et al.(2001)]{ham01} 
Hambly, N. C., Davenhall, A. C., Irwin, M. J., \& MacGillivray, H. T. 2001,
\mnras, 326, 1315

\bibitem[Hanson et al.(2004)]{han04} 
Hanson, R. B., Klemola, A. R., Jones, B. F., \& Monet, David G. 2004, \aj,
128, 1430

\bibitem[Harris et al.(1999)]{harr99} 
Harris, H. C., et al. 1999, \aj, 117, 339

\bibitem[Harris et al.(1988)]{har88} 
Harris, S., Clegg, P., \& Hughes, J. 1988, \mnras, 235, 441

\bibitem[Hartmann et al.(1999)]{har99} 
Hartmann, L., Calvet, N., Allen, L., Chen, H., \& Jayawardhana, R. 1999, 
\aj, 118, 1784

\bibitem[Hartmann et al.(1986)]{har86} 
Hartmann, L., Hewett, R., Stahler, S., \& Mathieu, R. D. 1986, \apj, 309, 275

\bibitem[Hartmann et al.(1987)]{har87} 
Hartmann, L. W., Soderblom, D. R., \& Stauffer, J. R. 1987, \aj, 93, 907

\bibitem[Hartmann et al.(2005b)]{har05} 
Hartmann, L., Megeath, S. T., Allen, L., Luhman, K., Calvet, N., 
D'Alessio, P., Franco-Hernandez, R., \& Fazio, G. 2005b, \apj, 629, 881

\bibitem[Hartmann et al.(2005a)]{har05st} 
Hartmann, L., et al. 2005a, \apj, 628, L147

\bibitem[Henry et al.(1994)]{hen94} 
Henry, T. J., Kirkpatrick, J. D., \& Simons, D. A. 1994, \aj, 108, 1437

\bibitem[Herbig(1977)]{her77} 
Herbig, G. H. 1977, \apj, 214, 747

\bibitem[Hillenbrand(1997)]{hil97}
Hillenbrand, L. A. 1997, \aj, 113, 1733

\bibitem[Hillenbrand \& Carpenter(2000)]{hc00} 
Hillenbrand, L. A., \& Carpenter, J. M. 2000, \apj, 540, 236

\bibitem[H{\o}g et al.(2000)]{hog00} 
H{\o}g, E., et al. 2000, \aap, 355, L27

\bibitem[Hughes et al.(2007)]{hug07} 
Hughes, A. M., Wilner, D. J., Calvet, N., D'Alessio, P., Claussen, M. J., 
\& Hogerheijde, M. R. 2007, \apj, 664, 536

\bibitem[Hughes et al.(2009)]{hug09} 
Hughes, A. M., et al. 2009, \apj, 698, 131

\bibitem[Jansen et al.(2001)]{jan01} 
Jansen, F., et al. 2001, \aap, 365, L1

\bibitem[Jones \& Herbig(1979)]{jh79} 
Jones, B. F., \& Herbig, G. H. 1979, \aj, 84, 1872

\bibitem[Kenyon et al.(1994)]{ken94} 
Kenyon, S. J., G\'omez, M., Marzke, R. O., \& Hartmann, L. 1994, \aj, 108, 251

\bibitem[Kenyon et al.(2008)]{ken08} 
Kenyon, S. J., G\'omez, M., \& Whitney, B. A. 2008, Handbook of Star 
Forming Regions, Volume 1, ASP Monograph Series, 405

\bibitem[Kerr \& Lynden-Bell(1986)]{ker86}
Kerr, F. J., \& Lynden-Bell, D. 1986, \mnras, 221, 1023

\bibitem[Kirkpatrick et al.(1997)]{kir97} 
Kirkpatrick, J. D., Henry, T. J., \& Irwin, M. J. 1997, \aj, 113, 1421

\bibitem[Kirkpatrick et al.(1991)]{kir91} 
Kirkpatrick, J. D., Henry, T. J., \& McCarthy, D. W. 1991, \apjs, 77, 417

\bibitem[Kirkpatrick et al.(1999)]{kir99} 
Kirkpatrick, J. D., et al. 1999, \apj, 519, 802

\bibitem[Kirkpatrick et al.(2006)]{kir06} 
Kirkpatrick, J. D., et al. 2006, \apj, 639, 1120

\bibitem[Kirkpatrick et al.(2008)]{kir08} 
Kirkpatrick, J. D., et al. 2008, \apj, 689, 1295

\bibitem[Kraus \& Hillenbrand(2007)]{kra07} 
Kraus, A. L., \& Hillenbrand, L. A. 2007, \apj, 662, 413

\bibitem[Lada(1987)]{lada87} 
Lada, C. J. 1987, in IAU Symp. 115, Star Forming Regions, ed. M. Peimbert 
\& J. Jugaku (Dordrecht: Reidel), 1

\bibitem[Lada et al.(2008)]{lada08} 
Lada, C. J., Muench, A. A., Rathborne, J., Alves, J. F., \& Lombardi, M. 
2008, \apj, 672, 410

\bibitem[Lawson et al.(2002)]{law02} 
Lawson, W. A., Crause, L. A., Mamajek, E. E., \& Feigelson, E. D. 2002, 
\mnras, 329, L29

\bibitem[Lee et al.(1999)]{lee99} 
Lee, C. W., Myers, P. C., \& Tafalla, M. 1999, \apj, 526, 788

\bibitem[Loinard et al.(2005)]{loi05} 
Loinard, L., Mioduszewski, A. J., Rodr{\'\i}guez, L. F., Gonz\'alez, R. A., 
Rodr{\'\i}guez, M. I., \& Torres, R. M. 2005, \apjl, 619, L179

\bibitem[Loinard et al.(2007)]{loi07} 
Loinard, L., et al. 2007, \apj, 671, 546

\bibitem[Luhman(1999)]{luh99} 
Luhman, K. L. 1999, \apj, 525, 466

\bibitem[Luhman(2004a)]{luh04cha} 
Luhman, K. L. 2004a, \apj, 602, 816

\bibitem[Luhman(2004b)]{luh04eta2} 
Luhman, K. L. 2004b, \apj, 616, 1033

\bibitem[Luhman(2004c)]{luh04tau} 
Luhman, K. L. 2004c, \apj, 617, 1216

\bibitem[Luhman(2006)]{luh06tau1} 
Luhman, K. L. 2006, \apj, 645, 676

\bibitem[Luhman(2007)]{luh07cha} 
Luhman, K. L. 2007, \apjs, 173, 104

\bibitem[Luhman et al.(2003a)]{luh03tau} 
Luhman, K. L., Brice\~{n}o, C., Stauffer, J. R., Hartmann, L., Barrado y 
Navascu\'{e}s, D., \& Nelson, C. 2003a, \apj, 590, 348

\bibitem[Luhman et al.(2009b)]{luh09tau} 
Luhman, K. L., Allen, P. R., Espaillat, C., Hartmann, L., \& Calvet, N. 2009b,
\apjs, submitted

\bibitem[Luhman et al.(2009a)]{luh09fu}
Luhman, K. L., Mamajek, E. E., Allen, P. R., Muench, A. A., \& 
Finkbeiner, D. P.  2009a, \apj, 691, 1265

\bibitem[Luhman \& Muench(2008)]{luh08cha2} 
Luhman, K. L., \& Muench, A. A. 2008, \apj, 684, 654

\bibitem[Luhman et al.(2003b)]{luh03ic} 
Luhman, K. L., Stauffer, J. R., Muench, A. A., Rieke, G. H., Lada, E. A., 
Bouvier, J., \& Lada, C. J. 2003b, \apj, 593, 1093

\bibitem[Luhman \& Steeghs(2004)]{luh04eta1} 
Luhman, K. L., \& Steeghs, D. 2004, \apj, 609, 917

\bibitem[Luhman et al.(2006)]{luh06tau2} 
Luhman, K. L., Whitney, B. A., Meade, M. R., Babler, B. L., Indebetouw, R., 
Bracker, S., \& Churchwell, E. B. 2006, \apj, 647, 1180

\bibitem[Luhman et al.(2008)]{luh08cha1} 
Luhman, K. L., et al. 2008, \apj, 675, 1375

\bibitem[Lyo et al.(2004)]{lyo04} 
Lyo, A.-R, Lawson, W. A., Feigelson, E. D., \& Crause, L. A. 2004, \mnras, 
347, 246

\bibitem[Lyo et al.(2006)]{lyo06} 
Lyo, A.-R, Song, I., Lawson, W. A., Bessell, M. S., \& Zuckerman, B. 2006, 
\mnras, 368, 1451

\bibitem[Malaroda et al.(2006)]{mal06} 
Malaroda, S., Levato, H., \& Galliani, S. 2006, VizieR Online Data Catalog, 
3249, 0

\bibitem[Mamajek et al.(1999)]{mam99} 
Mamajek, E., Lawson, W. A., \& Feigelson, E. D. 1999, \apj, 516, 77

\bibitem[Mart{\'{\i}}n et al.(2005)]{mar05} 
Mart{\'{\i}}n, E. L., Magazz{\`u}, A., Delfosse, X., \& Mathieu, R. D. 2005, 
\aap, 429, 939

\bibitem[Mathieu et al.(1997)]{mat97} 
Mathieu, R. D., Stassun, K., Basri, G., Jensen, E. L. N., Johns-Krull, C. M.,
Valenti, J. A., \& Hartmann, L. W. 1997, \aj, 113, 1841

\bibitem[Mohanty et al.(2005)]{moh05} 
Mohanty, S., Jayawardhana, R., \& Basri, G. 2005, \apj, 626, 498

\bibitem[Monet et al.(2003)]{mon03} 
Monet, D. G., et al. 2003, \aj, 125, 984

\bibitem[Moraux et al.(2007)]{mor07} 
Moraux, E., Lawson, W. A., \& Clarke, C. 2007, \aap, 473, 163

\bibitem[Muench et al.(2002)]{mue02} 
Muench, A. A., Lada, E. A., Lada, C. J., \& Alves, J. 2002, \apj, 573, 366

\bibitem[Muench et al.(2007)]{mue07} 
Muench, A. A., Lada, C. J., Luhman, K. L., Muzerolle, J., \& Young, E. 2007, 
\aj, 134, 411

\bibitem[Muench et al.(2003)]{mue03} 
Muench, A. A., et al. 2003, \aj, 125, 2029

\bibitem[Mundt et al.(1983)]{mun83} 
Mundt, R., Walter, F. M., Feigelson, E. D., Finkenzeller, U., Herbig, G. H., 
\& Odell, A. P. 1983, \apj, 269, 229

\bibitem[Muzerolle et al.(2003)]{muz03}
Muzerolle, J., Hillenbrand, L., Calvet, N., Brice\~{n}o, C., \& Hartmann, L.  
2003, \apj, 592, 266

\bibitem[Muzerolle et al.(2005)]{muz05} 
Muzerolle, J., Luhman, K. L., Brice\~{n}o, C., Hartmann, L., \& Calvet, N.  
2005, \apj, 625, 906

\bibitem[Onishi et al.(2002)]{oni02} 
Onishi, T., Mizuno, A., Kawamura, A., Tachihara, K., \& Fukui, Y. 2002, \apj, 
575, 950

\bibitem[Palla \& Stahler(1999)]{pal99}
Palla, F., \& Stahler, S. W. 1999, \apj, 525, 772

\bibitem[Prisinzano et al.(2008)]{pri08} 
Prisinzano, L., et al. 2008, \apj, 677, 401

\bibitem[Prusti et al.(1992)]{pru92} 
Prusti, T., Clark, F. O., Laureijs, R. J., Wakker, B. P., \& Wesselius, P. R.  
1992, \aap, 259, 537

\bibitem[Rayner et al.(2003)]{ray03}
Rayner, J. T., et al. 2003, \pasp, 115, 362

\bibitem[Reid \& Hawley(1999)]{rei99} 
Reid, I. N., \& Hawley, S. L. 1999, \aj, 117, 343

\bibitem[Reipurth et al.(1990)]{rei90} 
Reipurth, B., Lindgren, H., Nordstrom, B., \& Mayor, M. 1990, \aap, 235, 197

\bibitem[Rieke \& Lebofsky(1985)]{rl85}
Rieke, G. H., \& Lebofsky, M. J. 1985, \apj, 288, 618

\bibitem[Rieke et al.(2004)]{rie04} 
Rieke, G. H. et al. 2004, \apjs, 154, 25

\bibitem[R\"oser et al.(2008)]{ros08} 
R\"oser, S., Schilbach, E., Schwan, H., Kharchenko, N. V., Piskunov, A. E., \& 
Scholz, R.-D. 2008, \aap, 488, 401

\bibitem[Sartoretti et al.(1998)]{sar98} 
Sartoretti, P., Brown, R. A., Latham, D. W., \& Torres, G. 1998, \aap, 334, 592

\bibitem[Scelsi et al.(2007)]{sce07} 
Scelsi, L., et al. 2007, \aap, 468, 405

\bibitem[Scelsi et al.(2008)]{sce08} 
Scelsi, L., et al. 2008, \aap, 490, 601

\bibitem[Schmidt-Kaler(1982)]{sk82} 
Schmidt-Kaler, T. 1982, in Landolt-Bornstein, Group VI, Vol. 2, ed. K.-H.  
Hellwege (Berlin: Springer), 454

\bibitem[Skrutskie et al.(2006)]{skr06} 
Skrutskie, M., et al. 2006, \aj, 131, 1163

\bibitem[Slesnick et al.(2006)]{sle06} 
Slesnick, C. L., Carpenter, J. M., Hillenbrand, L. A., \& Mamajek, E. E.  
2006, \aj, 132, 2665

\bibitem[Song et al.(2004)]{sz04} 
Song, I., Zuckerman, B., \& Bessell, M. S. 2004, \apj, 600, 1016

\bibitem[Tatematsu et al.(2004)]{tat04} 
Tatematsu, K., Umemoto, T., Kandori, R., \& Sekimoto, Y. 2004, \apj, 606, 333

\bibitem[Telleschi et al.(2007)]{tel07} 
Telleschi, A., G\"udel, M., Briggs, K. R., Audard, M., \& Palla, F. 2007, \aap,
468, 425

\bibitem[Torres et al.(1995)]{tor95} 
Torres, C. A. O., Quast, G., de La Reza, R., Gregorio-Hetem, J., \& Lepine, 
J. R. D. 1995, \aj, 109, 2146

\bibitem[Torres et al.(2000)]{tor00} 
Torres, C. A. O., da Silva, L., Quast, G. R., de La Reza, R., \& Jilinksi, E. 
2000, \aj, 120, 1410

\bibitem[Torres et al.(2007)]{tor07} 
Torres, R. M., Loinard, L., Mioduszewski, A. J., \& Rodr{\'\i}guez, L. F. 2007, 
\apj, 671, 1813

\bibitem[Torres et al.(2009)]{tor09} 
Torres, R. M., Loinard, L., Mioduszewski, A. J., \& Rodr{\'\i}guez, L. F. 2009, 
\apj, 698, 242

\bibitem[Ungerechts \& Thaddeus(1987)]{ung87} 
Ungerechts, H., \& Thaddeus, P. 1987, \apjs, 63, 645

\bibitem[Vacca et al.(2003)]{vac03} 
Vacca, W. D., Cushing, M. C., \& Rayner J. T., 2003, \pasp, 115, 389

\bibitem[van Leeuwen(2007)]{van07} 
van Leeuwen, F. 2007, \aap, 474, 653

\bibitem[Walter(1986)]{wal86} 
Walter, F. M. 1986, \apj, 306, 573

\bibitem[Walter et al.(1988)]{wal88} 
Walter, F. M., Brown, A., Mathieu, R. D., Myers, P. C., \& Vrba, F. J. 1988, 
\aj, 96, 297

\bibitem[Walter et al.(2003)]{wal03} 
Walter, F. M., Beck, T. L., Morse, J. A., \& Wolk, S. J. 2003, \aj, 125, 2123

\bibitem[Watson \& Stapelfeldt(2007)]{wat07} 
Watson, A. M., \& Stapelfeldt, K. R. 2007, \aj, 133, 845

\bibitem[Werner et al.(2004)]{wer04} 
Werner, M. W., et al. 2004, \apjs, 154, 1

\bibitem[White \& Basri(2003)]{whi03} 
White, R. J., \& Basri, G. 2003, \apj, 582, 1109

\bibitem[White \& Ghez(2001)]{wg01} 
White, R. J., \& Ghez, A. M. 2001, \apj, 556, 265

\bibitem[White \& Hillenbrand(2004)]{whi04} 
White, R. J., \& Hillenbrand, L. A. 2004, \apj, 616, 998

\bibitem[Wichmann et al.(1998)]{wic98} 
Wichmann, R., Bastian, U., Krautter, J., Jankovics, I., \& Ruci\'nski, S. M.  
1998, \mnras, 301, L39

\bibitem[Zacharias et al.(2004a)]{zac04a} 
Zacharias, N., Monet, D. G., Levine, S. E., Urban, S. E., Gaume, R., \& Wycoff, 
G. L. 2004a, \baas, 36, 4815

\bibitem[Zacharias et al.(2004b)]{zac04b} 
Zacharias, N., Urban, S. E., Zacharias, M. I., Wycoff, G. L., Hall, D. M., 
Monet, D. G., \& Rafferty, T. J., 2004b, \aj, 127, 3043

\bibitem[Zaitseva et al.(1985)]{zai85} 
Zaitseva, G. V., Kolotilov, E. A., Petrov, P. P., Tarasov, A. E., Shenavrin, V. 
I., \& Shcherbakov, A. G. 1985, Soviet Astronomy Letters, 11, 109

\bibitem[Zaitseva et al.(1990)]{zai90} 
Zaitseva, G. V., Shcherbakov, A. G., \& Stepanova, N. A. 1990, Soviet Astronomy 
Letters, 16, 350

\bibitem[Zuckerman \& Song(2004)]{zuc04} 
Zuckerman, B., \& Song, I. 2004, \araa, 42, 685

\end{thebibliography}
\end{document}